\documentclass[11pt,leqno]{article}

\usepackage{amsthm,amsmath,natbib, mathtools}
\usepackage{ifthen}
\usepackage{enumitem}
\usepackage{amssymb,dsfont,url,color,booktabs}
\usepackage[x11names]{xcolor}
 \usepackage{tikz}
 \usepackage{geometry}
\usepackage{rotating}
 \usepackage{enumitem}

\usepackage{epstopdf}
\usepackage{chngcntr}
\usepackage{bbm}
\usepackage{array}
\usepackage[export]{adjustbox}[2011/08/13] 
\usepackage{caption}
\usepackage{multirow}

\usepackage[hang]{subfigure} 
\usepackage{wrapfig} 
\usepackage{tikz}
\usetikzlibrary{arrows, patterns}
\tikzset{
  schraffiert/.style={pattern=horizontal lines,pattern color=#1},
  schraffiert/.default=black
}
\usepackage{todonotes}

\newcommand*{\IR}{\mathds{R}}

\newcommand*{\IC}{\mathds{C}}

\newcommand*{\BS}{\text{Black-Scholes }}

\renewcommand*\d{\mathop{}\!\mathrm{d}}

\theoremstyle{plain}
\newtheorem{theorem}{Theorem}[section]
\newtheorem{lemma}[theorem]{Lemma}
\newtheorem{proposition}[theorem]{Proposition}

\newtheorem{remark}[theorem]{Remark}

\newtheorem{assumptions}[theorem]{Assumptions}

\newcommand{\dd}[1]{\operatorname{d}\!#1}
\newcommand{\ccd}{\mathds{C}^d}
\newcommand{\ee}[1]{\operatorname{e}^{#1}}

\newcommand{\OF}{\mathcal{F}}
\newcommand{\OI}{\mathcal{I}}
\newcommand{\II}{\mathcal{I}}
\newcommand{\OK}{\mathcal{K}}
\newcommand{\OP}{\mathcal{P}}
\newcommand{\OQ}{\mathcal{Q}}

\newcommand{\OT}{\mathcal{T}}
\newcommand{\OU}{\mathcal{U}}
\newcommand{\OH}{\mathcal{U}}
\newcommand{\OX}{\mathcal{X}}
\newcommand{\OEE}{\mathcal{E}}
\newcommand{\notiz}[1]{\relax}

\newcommand{\zitep}[1]{\relax}
\newcommand{\skr}{\rangle}
\newcommand{\1}{\mathds 1}            

\newcommand{\nn}{\mathds N}

\newcommand{\rr}{\mathds R}

\newcommand{\rrd}{\mathds{R}^d}

\newcommand{\cc}{\mathds C}
\newcommand{\skl}{\langle}

\newcommand{\argmax}{\operatorname{arg\,max}}

\newcommand{\modelp}{q}
\newcommand{\modelP}{\OQ}

\newcommand{\Price}[1][]{
		\ifthenelse{\equal{#1}{}}{\mathit{Price}}{\Price{}^{#1}}
	} 
\newcommand{\Pricekp}{\Price^{K,T,q}}
\usepackage{mathrsfs}

\newcommand{\ulb}{\underline{b}}	
\newcommand{\olb}{\overline{b}}

\newcommand{\tild}{~}

\newlength{\wordlength}

\begin{document}
\title{Magic Points in Finance: \\
Empirical Integration for 
Parametric  Option Pricing\footnote{We gratefully acknowledge Bernard Haasdonk, Laura Iapichino,  Yvon Maday, Anthony Patera and Barbara Wohlmuth for fruitful discussions on the magic point empirical interpolation. 
 Moreover, we like to thank Ernst Eberlein, Sergei Levendorski\u{i} and Dilip Madan for valuable feedback. 
Additionally, we thank the participants of the conferences MoRePas–2015: Model Reduction of Parametrized Systems III, held in Trieste, Vienna Congress on Mathematical Finance -- VCMF 2016, Vienna. September 12--14, 2016, and the 9th World Congress of the Bachelier Finance Society, New York. July 15--19, 2016. 
Furthermore we thank the participants of the research seminars 
Bachelier Seminaire, Paris. January 15, 2016,
Mathematics Institute of Computational Science and Engineering MATHICSE, EPFL Lausanne. January 13, 2016.
Maximilian Ga{\ss} thanks the KPMG Center of Excellence in Risk Management  and Kathrin Glau acknowledges the TUM Junior Fellow Fund for financial support.
}} 

\bigskip
\author{\textbf{Maximilian Ga{\ss}, Kathrin Glau, Maximilian Mair
}
\\\\Department of Mathematics,\\ Technical University of Munich, Germany\\\indent  kathrin.glau@tum.de
}

\date{\today, first version: November, 3, 2015}
\maketitle



\begin{abstract}
We propose an offline-online procedure for Fourier transform based option pricing. The method supports the acceleration of such essential tasks of mathematical finance as model calibration, real-time pricing, and, more generally, risk assessment and parameter risk estimation. 
We adapt the empirical magic point interpolation method of \ Barrault, Nguyen, Maday and Patera (2004)\nocite{BarraultNguyenMadayPatera2004} to parametric Fourier pricing. 
In the offline phase, a quadrature rule is tailored to the family of integrands of the parametric pricing problem. In the online phase, the quadrature rule then yields fast and accurate approximations of the option prices. Under analyticity assumptions the pricing error decays exponentially.
Numerical experiments in one dimension confirm our theoretical findings and show a significant gain in efficiency, even for examples beyond the scope of the theoretical results.
\end{abstract}

	




\newpage

\section{Introduction}

Most of the option pricing methods based on Fourier transforms aim at the evaluation of individual option prices. For this use case, existing pricing tools have achieved impressive performance. For real-time applications and those involving repeated evaluations particularly fast run-times are crucial. Therefore, Fast Fourier transforms (FFT) have become highly popular to reduce computational complexity when prices are required simultaneously for a large set of different strikes, following the seminal works of \cite{CarrMadan99} and \cite{Raible}. See also the monograph \cite{BoyarchenkoLevendorskii2002}.
In this paper we shift the focus from the pricing problem for one or several strikes to the full \textit{parametric option pricing problem}, considering \textit{all parameters} such as strike, maturity, and model parameters.

For a given model and option type, various applications require the evaluation of Fourier pricing routines repeatedly for different parameter constellations. We mention three of those applications: Firstly, during calibration of financial models with Fourier methods, the optimization relies on multiple evaluations of a Fourier integral for varying model and option parameters. Secondly, each (intra-)day recalibration leads to new model parameters for which several option prices and their sensitivities have to be computed in real-time. Thirdly, a variety of other relevant financial quantities that need to be repeatedly evaluated for different parameter constellations can be expressed via Fourier transforms analogously to the Fourier representation of option prices. As one example we refer to \cite{ArmentiCrepeyDrapeauPapapantoleon2015} who propose a method to quantify risk allocation. At the heart of their algorithm lies an optimization routine that requires the repeated evaluation of parametric expectations that can be expressed as Fourier integrals. They give numerical examples, where standard Fourier methods turn out to be too slow for practical use. 

In all of these cases, Fourier pricing routines that are \textit{fast and accurate for a whole set of relevant parameter constellations} are required. Often, numerical experiments on the performance of pricing routines are presented for some fixed parameter constellations, see for instance \cite{Sydow2015}. Such comparisons clearly present highly valuable insight in different methods. It has, however, to be recognized that a method that performs highly efficient for some specific parameters needs not to be as efficient for the whole set of relevant parameters. This phenomenon has already been observed and covered by experiments, see \cite{InnocentisLevendorskiy2014}. In this article we present a new Fourier pricing method that is designed to be highly efficient for a whole range of parameters of interest.

We look at this Parametric Option Pricing (POP) through the lense of offline-online schemes. The main idea is to achieve fast and accurate real-time pricing, founded on a pre-computation step. The architecture of such methods decomposes into two separate phases. In the so-called \emph{offline phase}, the algorithm parses the complexity of the parametric pricing problem and extracts a structure bearing all of the important information on the whole problem as such. This is the computationally demanding part. Ideally, the offline phase is only performed once for a selected model class and option type. Thus the offline phase can be seen as part of the implementation of the pricing method. Intuitively, it represents the \textit{learning phase of the algorithm}. In the so-called \emph{online phase}, real-time pricing is performed. This second part benefits from the pre-computation and thus yields the desired fast and accurate pricing results.

Two types of offline-online schemes have been proposed for POP in the literature:
In \cite{SachsSchu2010}, \cite{ContLantosPironneau2011}, \cite{Pironneau2011} and \cite{HaasdonkSalomonWohlmuth2012b} offline-online decomposition has been adopted to solve parametric partial differential equations for option pricing.
 In contrast, \cite{GassGlauMahlstedtMair2015} present polynomial interpolation of the option price in the parameter space. 

In this paper, we tailor an offline-online scheme to Fourier pricing. Over the last fifteen years, Fourier based option pricing has been applied successfully in both academia and practice. Pioneered by \cite{SteinStein1991} and \cite{Heston1993} for Brownian models, researchers have exploited the flexibility of the approach to create fast and efficient pricing algorithms for a large class of models and option types. (Fast) Fourier pricing of European options in L\'evy and the large class of affine models has first been developed by \cite{CarrMadan99}, \cite{Raible} and \cite{DuffiePanSingleton2000}. There is also a large and growing literature on Fourier methods to price path dependent options, see e.g.\ 
 \cite{BoyarchenkoLevendorskiy2002b}, \cite{FengLinetsky2008}, \cite{KudryavtsevLevendorskiy2009}, \cite{Zhylyevskyy2010},  \cite{FangOosterlee2011}, \cite{LevendorskiyXie2012}, \cite{FengLin2013} and \cite{ZengKwok2014},
 and see \cite{EberleinGlauPapapantoleon2010} for a general framework and analysis.

The main contributions of this article are to
\begin{enumerate}
\item[--] propose an empirical quadrature rule to efficiently evaluate Fourier integrals for option pricing, 
\item[--]
find exponential convergence of the pricing error when the option price satisfies strict analyticity assumptions,
\item[--]
empirically observe exponential convergence of the method, even for examples beyond the scope of the theoretical results,
\item[--] empirically compare the efficiency of our method to the cosine method of \cite{FangOosterlee2008}.
\end{enumerate}

Our numerical and theoretical results show that the offline-online decomposition can be used to find a quadrature rule that offers very satisfying results in terms of accuracy and efficiency. As a further advantage, the offline-online scheme is build in such a way that the resulting quadrature rule satisfies a pre-specified accuracy for the whole parameter domain of interest.
Our univariate results lay the cornerstone for further research and show the potential for extensions of the method especially with regard to higher dimensions.

Parametric integrals also naturally arise in many other disciplines of applied mathematics. We refer to \cite{GassGlau2015} for this more general focus.
%

To achieve our goals, we apply the Empirical Magic Point Interpolation method developed by \cite{BarraultNguyenMadayPatera2004} in the context of  parametric nonlinear partial differential equations. 
While they enforce an affine decomposition of parametric operators, we decompose parametric Fourier integrands. Sketching our idea, the starting point is the Fourier representation of the parametric option price,
\begin{equation*}
\Pricekp=\frac{1}{(2\pi)^d} \int_{\Omega}  \widehat{f_{K}}(-z) \varphi_{T,q}(z) \dd z,
\end{equation*}
with generalized Fourier transform $\widehat{f_{K}}$ of the payoff function $f_K$ and the generalized Fourier transform $\varphi_{T,q}$ of the modelling random variable~$X_T^q$.
We follow the iterative Empirical Interpolation procedure outlined in \cite{MadayNguyenPateraPau2009}, that we describe in detail in section \ref{sec-magic} below.
For $M\in \nn$ the method recursively gives \textit{magic points} $z^\ast_1,\ldots,z^\ast_M \in \Omega$, \textit{basis functions} $q_1,\ldots,q_M$ and 
 $\theta_m^M := \sum_{j=1}^M(B^M)^{-1}_{jm} q_j$ and $B^M_{jm}:=q_m(z^\ast_j)$. The resulting price approximation is of the form
\begin{align}\label{approx-Pip}
\Pricekp\cong \frac{1}{(2\pi)^d} \sum_{m=1}^M \widehat{f_{K}}(-z^\ast_m) \varphi_{T,q}(z^\ast_m) \int_\Omega \theta_m^M(z) \dd z.
\end{align}

The algorithm naturally decomposes into two phases. An offline phase, where the just mentioned quantities are constructed, and an online phase where real-time pricing is performed. More precisely, the two phases are described as follows.\vspace{1ex}

\noindent Offline phase: For a given parameter space,
\begin{itemize}
\item[--] identify the magic points $z^\ast_1,\ldots,z^\ast_M \in \Omega$, and
\item[--] precompute the integrals $\int_\Omega \theta_m^M(z) \dd z$ for all $m\le M$.
\end{itemize}
Online phase: For an arbitrary parameter constellation $(K,T,q)$,
\begin{itemize}
\item[--] evaluate the 
Fourier integrands $\widehat{f_{K}}(-z^\ast_m) \varphi_{T,q}(z^\ast_m)$ for all $m\le M$ and
\item[--] assemble the sum in \eqref{approx-Pip}.
\end{itemize}
In the cases we consider, the number of summands $M$ ranges in the dozens for a high accuracy already. Thus the evaluation of prices by\tild \eqref{approx-Pip} is fast and accurate.
The following features of our problems at hand are key for the efficiency of the online phase: Typically the mapping
\begin{equation*}
(K,T,q,z)\mapsto\widehat{f_{K}}(-z) \varphi_{T,q}(z),
\end{equation*}
i.e. the parametric integrand in \eqref{approx-Pip}
\begin{itemize}
\item[(i)] 
is \emph{explicitly available}, and 
\item[(ii)] 
enjoys desirable \emph{analyticity properties}.
 \end{itemize}
Thanks to (i), the evaluation of a single summand in \eqref{approx-Pip} is effortless, and thanks to (ii), a few summands already yield high accuracy.
  In our numerical experiments for option pricing in univariate models, we achieve average absolute pricing accuracies ranging from $10^{-6}$ to $10^{-10}$ for $40$ to $50$ magic points, depending on the model used.
 
This article is organized as follows: In the next section we revisit the framework for Fourier pricing in detail. In section \ref{sec-magic}  we adapt the Empirical Magic Point Interpolation method to Fourier pricing and describe the resulting algorithm that we call MagicFT. Based on Theorem 2.4 in \cite{MadayNguyenPateraPau2009}, we present in section \ref{sec-convergence} \emph{exponential convergence results} under suitable analyticity conditions along with explicit error bounds. We investigate these analyticity properties for different payoff profiles and models in section \ref{sec-Ex+Case}. In section \ref{sec-num} we implement the algorithm and perform an empirical convergence study. In several case studies we investigate the MagicFT approximation for several models individually. Moreover we compare the MagicFT method to the popular cosine method of \cite{FangOosterlee2008}. 
We conclude with an appendix that highlights essential features of  Empirical Magic Point Interpolation and, for the sake of self-contained readability, presents detailed proofs of the convergence results.

\section{Parametric Option Pricing with Fourier Transform}\label{sec-FourierPrice}
We compute option prices of the form 
\begin{equation}\label{eq-price}
\Pricekp:= E \big[ f_K(X^q_T)\big]
\end{equation}
with parametrized payoff function $f_K:\rrd\to \rr$ and a parametric $\OF_T$-measurable $\rrd$-valued random variable $X^q_T$. Here we use \textit{payoff and model parameters} $K\in\OK\subset\rr^{D_1}$, $T\in\OT\subset\rr^{D_2}$, $q\in\OQ\subset\rr^{D_3}$ and denote $D=D_1+D_2+D_3$. Furthermore, let
\begin{equation*}
p=(K,T,q)\in \OP\quad\text{where }\OP=\OK\times\OT\times\OQ.
\end{equation*}
 
In order to pass to the pricing formula in terms of Fourier transforms, we impose the following \emph{exponential moment condition} for $\eta\in \rrd$,
\begin{align}\label{Expeta}
E\big[ \ee{- \skl\eta ,X^q_T\skr }\big] < \infty\quad\text{for all }(T,q)\in\OT\times\OQ,\tag{Exp}
\end{align}
which allows us to define for every $(T,q)\in\OT\times\OQ$ the extension of the characteristic function of $X_T^q$ to the complex domain $\rrd + i\eta$,
\begin{align}\label{phip}
\varphi_{T,q}(z):=E\big[ \ee{i \skl z ,X^q_T\skr }\big] \qquad \text{for all }z=\xi + i\eta, \,\xi \in \rrd.
\end{align}
We further introduce the following integrability condition 
\begin{align}\label{Int}
\scalebox{.98}[1]{$\displaystyle  x\mapsto \ee{\skl \eta, x\skr}\!f_K(x), \, \xi \mapsto \varphi_{T,q}(\xi + i\eta) \in L^1(\rrd) \,\text{for all }(K,T,q)\in\OP.$}\tag{Int}
\end{align}
Furthermore, we denote
\begin{align}
\widehat{f_K}(\xi + i\eta):= \int_{\rrd} \ee{i\skl\xi +i\eta,x \skr} f_K(x)\dd x,
\end{align}
the \emph{generalized Fourier transform} of $f_K$.
The Fourier representation of option prices traces back to the pioneering works of \cite{CarrMadan99} and \cite{Raible}. The following version is an immediate consequence of Theorem 3.2 in \cite{EberleinGlauPapapantoleon2010}.
\begin{proposition}[Fourier pricing]\label{prop-FT}
Let $\eta\in \rrd$ such that \emph{(Exp)} and \emph{(Int)} are satisfied. Then for every $(K,T,q)\in\OP$,
\begin{align}\label{Fourierprice}
\Pricekp = \frac{1}{(2\pi)^d} \int_{\rrd + i\eta }  \widehat{f_K}(-z) \varphi_{T,q}(z) \dd z.
\end{align}
\end{proposition} 
Typically, that is for the most common option types, the generalized Fourier transform of $f_K$ is of the form 
\begin{equation}
\widehat{f_K}(z) = K^{iz+c}F(z)
\end{equation}
 for every $z\in\rrd + i\eta$ with some constant $c\in\rr$ and a function $F:\rrd + i\eta\to\cc$. Then the option prices \eqref{Fourierprice} are indeed parametric Fourier integrals of the form
 \begin{align}\label{parametricFourierprice}
\Pricekp = \frac{1}{(2\pi)^d} \int_{\rrd + i\eta } \ee{-i\skl z,\log(K)\skr} K^cF(z)\varphi_{T,q}(z) \dd z.
\end{align}
%
As a first step in the numerical evaluation of \eqref{parametricFourierprice} we
 employ an elementary symmetry by virtue of the identity $\widehat{f}(-\xi) = \overline{\widehat{f}(\xi)}$ for all $\xi\in\rrd$ and all real-valued integrable functions $f$, and obtain
\begin{align}\label{eq-FT-realpart}
\int_{\rr^{d}+ i\eta}
\widehat{f_K}(-z) \varphi_{T,q}(z) \dd z = 2 
\int_{\rr_+\times\rr^{d-1}+i\eta}
\Re\Big(\widehat{f_K}(-z) \varphi_{T,q}(z)\Big) \dd z,
\end{align}
which reduces the numerical effort by half.

In a second step we restrict the domain of integration to a compact set $\Omega\subset \rrd$. The resulting error is determined by the decay of the integrand and will be further analyzed in appendix~\ref{sec:MFTTruncationError}.
From now on we set $\Omega:=\Omega_1\times\ldots \times \Omega_d$ with bounded open intervals $\Omega_1\subset \rr^+ + i\eta_1$ and $\Omega_j\subset \rr+i\eta_j$ for $j=2,\ldots,d$. 

\section{Magic Point Interpolation for Integration}\label{sec-magic}

We present the \emph{Empirical Magic Point Interpolation method for parametric integration} to approximate parametric integrals of the form
\begin{align}\label{parametric Integration}
\OI(h_p) :=  \int_{\Omega}  h_p(z) \dd z\qquad\text{for }p\in \OP
\end{align}
with the parametric integrands
\begin{align}
h_{p}(z)  = h_{(K,T,q)}(z)  &:= \Re\big(\widehat{f_K}(-z) \varphi_{T,q}(z)\big)\label{Def-h}
\end{align}
for every $p=(K,T,q)$ in a given parameter set $ \OP$. With $\OP$ we associate
\begin{align}
\OU&:= \big\{ h_{p}:\Omega\to \rr\,|\, p \in \OP \}, \label{Def-Uallg}
\end{align}
the set of all parametric integrands.
%
Let us point out that the following iterative procedure is defined for a more general set of parametric integrands that are not required to be of the form \eqref{Def-h}.

Before we closely follow \cite{MadayNguyenPateraPau2009} to describe the interpolation method let us state our basic assumptions that ensure the well-definedness of the iterative procedure. 

\begin{assumptions} \label{assumption:magic}
Let $\left(\Omega,\|.\|_{\infty}\right)$ and $\left(\OP,\|.\|_{\infty}\right)$ be compact, $\OP\times \Omega\ni (p,z)\mapsto h_p(z)$ bounded and $p\mapsto h_p$ be sequentially continuous, i.e. for every sequence $p_i\to p$ we have  $\|h_{p_i}-h_{p}\|_\infty\to0$. Moreover, $\OU$ is nontrivial in the sense that the set contains elements other than the function that is constantly zero. 
\end{assumptions}

For $M\in \nn$ define a mapping $I_M$ from $\OU$ to a tensor specified by
\begin{align}
I_M(h)(p,z):= \sum_{m=1}^M h_{p}(z^\ast_m) \theta_m^M(z)
\end{align}
and the \emph{Magic Point Integration} with $M$ points by
\begin{align}\label{Int_M}
\II_M(h)(p):= \sum_{m=1}^M h_{p}(z^\ast_m) \int_{\Omega} \theta_m^M(z) \dd z
\end{align}
with
\begin{align}\label{def-theta}
\theta_m^M(z) := \sum_{j=1}^M(B^M)^{-1}_{jm} q_j(z), \qquad 
B^M_{jm}:=q_m(z^\ast_j),
\end{align}
where we denote by $(B^M)^{-1}_{jm}$ the entry in the $j$th line and $m$th column of the inverse of matrix $B^M$. By definition, $B^M$ is a lower triangular matrix with unity diagonal and is thus invertible, confer also section~\ref{sec:MFTGeneralFeatures} in the appendix. The \textit{magic points} $z^\ast_1,\ldots,z^\ast_M \in \Omega$ and the \textit{basis functions} $q_1,\ldots,q_M$ are recursively defined in the following way:

In the first step, let
\begin{align}\label{def_u1}
u_1 := \underset{u\in \OU}{\argmax}\|u\|_{\infty},\quad z^\ast_1:=\underset{z\in \Omega}{\argmax}|u_1(z)|,\quad q_1(\cdot):=\frac{u_1(\cdot)}{u_1(z^\ast_1)}.
\end{align}
Note that thanks to Assumption \ref{assumption:magic}, these operations are well-defined.
Then, recursively, as long as there are at least $M$ linearly independent functions in\tild $\OU$, $u_M$ is chosen according to a greedy procedure: The algorithm chooses $u_M$ as the function in the set $\mathcal{U}$ which is worst represented by the approximation with the previously identified $M-1$ magic points and basis functions, 
\begin{align}\label{defu_M}
u_M := \underset{u\in \OU}{\argmax}\|u - I_{M-1}(u)\|_{\infty}.
\end{align}
Since every $u\in\mathcal{U}$ is a parametric function, $u=h_p$ for some $p\in\mathcal{P}$, it can be identified by the associated parameter $p$. We call $p_M^\ast\in\mathcal{P}$ identifying $u_M$ in\tild \eqref{defu_M} the $M$th \emph{magic parameter}. In the same spirit,\tild let
\begin{align}\label{defxi_M}
z^\ast_M:=\underset{z\in \Omega}{\argmax}\big|u_M(z) - I_{M-1}(u_M)(z)\big|,
\end{align}
and we call $z^\ast_M$ the $M$th \emph{magic point}. The $M$th \emph{basis function} is the residual, normed to $1$, when evaluated at the new magic point $z^\ast_M$,
\begin{align}\label{def-qM}
 q_M(\cdot):=\frac{u_M(\cdot) - I_{M-1}(u_M)(\cdot) }{u_M(z^\ast_M) - I_{M-1}(u_M)(z^\ast_M)}.
\end{align}
Note the well-definedness of the operations in the iterative step thanks to Assumption \ref{assumption:magic} and the fact that the denominator in \eqref{def-qM} is only zero, if all functions in $\OU$ are perfectly represented by the interpolation $I_{M-1}$, in which case they span a linear space of dimension $M-1$ or less and the procedure would have stopped already.

We may take three different perspectives on the approach:
\begin{enumerate}[label=(\roman*), leftmargin=3em]
\item
Magic Point Integration is a \emph{quadrature rule for integrating parametric functions}, where the interpolation nodes are chosen in a precomputation phase according to the set of integrands at hand.
\item
Consider that for $m=1,\ldots,M$ the functions $\theta^M_m$ are linear combinations of snapshot integrands $h_{p^\ast_m}$ with coefficients $\beta^m_1,\ldots,\beta^m_M$ and hence
\begin{align}\label{magic-interpolation-integral}
\OI_M(h)(p) = \sum_{m=1}^M   h_{p}(z^\ast_m) \sum_{j=1}^M \beta^m_j \int_\Omega h_{p_j^\ast}(z)\dd z.
\end{align}
This means that Magic Point Integration is an  \emph{interpolation method for parametric integrals} in the parameter space.
Thus, taking the error stemming from truncating the integration domain to $\Omega$ into account, equation \eqref{magic-interpolation-integral} induces an approximation of the option price by a linear combination of snapshot prices,
\begin{align}\label{magic-interpolation-price}
\Pricekp\cong\sum_{m=1}^M   h_{(K,T,q)}(z^\ast_m) \sum_{j=1}^M \beta^m_j \Price^{K_j,T_j,q_j}.
\end{align}
\item In view of the representation of the option prices $\Pricekp$ as parametric Fourier integrals in \eqref{parametricFourierprice}, we use  the Magic Point Integration algorithm to approximate parametric Fourier transforms that we call \emph{MagicFT} as introduced in \cite{GassGlau2015}.
\end{enumerate}

From perspective (i), Magic Point Integration for parametric option pricing is an alternative to standard quadrature rules. Standard integration routines suffer from the curse of dimensionality of the integration domain. In contrast, under suitable analyticity conditions, the approximation error of Magic Point Integration decays exponentially in $M$, independently of the dimension of $\Omega$, if the parameter space is one-dimensional, see Theorem \ref{theo:MFTExpConvEU} below.

 Taking the point of view (ii), Magic Point Integration for parametric option pricing can be compared to a benchmark method for parametric option pricing by interpolation.  Standard interpolation methods in the parameter suffer from the curse of dimensionality of the parameter space. In contrast, under suitable analyticity conditions on the integrands, the approximation error of Magic Point Integration decays exponentially in $M$, independently of the dimension of $\OP$, if the integration domain is one-dimensional, see Theorem \ref{theo:MFTExpConvBasket} below.

\section{Convergence Analysis of Magic Point Integration}\label{sec-convergence}

To show the virtue of the method in its full generality, we review a general convergence result for Magic Point Interpolation derived in \cite{MadayNguyenPateraPau2009}. This result relates the convergence of Magic Point Interpolation to the best linear $n$-term approximation that is formally expressed by the Kolmogorov $n$-width. For a real or complex normed linear space $\big(\OX,\|\cdot\|\big)$ and $\OU\subset\OX$, the \emph{Kolmogorov $n$-width} is given by
\begin{equation}
d_n(\OU,\mathcal{X})=\inf_{\OU_n\in\OEE(\OX,n)}\sup_{g\in\mathcal{U}}\inf_{f\in\OU_n}\|g-f\|,\label{eq:nwidth}
\end{equation}
where $\OEE(\OX,n)$ is the set of all $n$ dimensional subspaces of $\OX$.
%

We denote by $\big(L^\infty(\Omega,\cc),\|\cdot\|_{\infty}\big)$ the Banach space of functions mapping from $\Omega\subset \ccd$ to $\cc$ that are bounded in the supremum norm.

%


\begin{proposition}\label{prop-conv}
For the set $\OU$ from \eqref{Def-Uallg} and $M\in\nn$
\begin{enumerate}[label=$(A\arabic*)$,leftmargin=2.5em]
\item
assume $\Omega\subset\cc^d$ and Assumption \ref{assumption:magic},

\item
assume there exist constants $\alpha>\log(4)$ and $c>0$ such that
\[
d_M\big(\OU, L^\infty(\Omega,\cc) \big)\le c\ee{-\alpha M}.
\]
\end{enumerate}
Then for arbitrary $\varepsilon>0$  and $C:=\frac{c}{4}\ee{\alpha}+\varepsilon$ we have for all $u\in\mathcal{U}$ that
\begin{equation}
\big\|u-I_M(u)\big\|_{\infty}\le CM\ee{-(\alpha-\log(4))M}.
\end{equation}
\end{proposition}
The proposition directly follows from Theorem 2.4 in \cite{MadayNguyenPateraPau2009}, where a slightly different version that does not explicitly use the Kolmogorov $n$-width is provided. In order to keep our presentation self-contained and as transparent as possible, we present a detailed proof of the Proposition in Appendix \ref{sec-insight-magic}, where we also highlight essential features of the iterative Magic Point Interpolation procedure.

%

\subsection{Exponential Convergence of Magic Point Integration for Parametric Option Pricing}

In order to formulate our analyticity assumptions, we define the 
\textit{Bernstein ellipse} $B([-1,1],\varrho)$ with parameter $\varrho>1$ as the open region in the complex plane bounded by the ellipse with foci $\pm 1$ and semiminor and semimajor axis lengths summing up to $\varrho$ with the origin as the center and semimajor axis on the real axis. Moreover, we define for $\ulb<\olb\in\rr$ the \textit{generalized Bernstein ellipse} by 
\begin{equation}
B([\ulb,\olb],\varrho):=\tau_{[\ulb,\olb]}\circ B([-1,1],\varrho),
\end{equation}
 where the transform $\tau_{[\ulb,\olb]}:\cc\to\cc$ is given by $\tau_{[\ulb,\olb]}\big(\Re(x)\big):=\olb + \frac{\ulb-\olb}{2}\big(1-\Re(x)\big)$ and $\tau_{[\ulb,\olb]}\big(\Im(x)\big):= \frac{\olb-\ulb}{2}\Im(x)$ for  every $x\in\cc$.

For an arbitrary set $X\subset\rr$, we define the generalized Bernstein ellipse by
\begin{equation}
B(X,\varrho) := B([\inf{X},\sup{X}],\varrho).
\end{equation} 

In order to estimate the error resulting from the Magic Point Interpolation method, we formulate two analyticity conditions. Condition\tild \ref{cond:MFTExpConvEU} is tailored to the case of univariate integration domains and\tild \ref{cond:MFTExpConvBasket} to the case of univariate parameter spaces.
\begin{enumerate}[label=$(B\arabic*)$,leftmargin=2.5em]
\item
\label{cond:MFTExpConvEU}
The function $(p,z)\mapsto h_p(z)$ is continuous on $\OP\times\Omega$ and there exist functions $H_1:\OP\times\Omega\to\cc$ and $H_2:\OP\to\cc$ such that for all $(p,z)\in \OP\times\Omega$,
\begin{equation*}
h_{p}(z) = H_1(p,z) H_2(p) 
\end{equation*}
and $H_1(p,z)$ has an extension $H_1:\OP\times B(\Omega,\varrho)\to\cc$ such that, for all fixed $p\in \OP$ the mapping $z\mapsto H_1(p,z)$ is analytic in the interior of the generalized Bernstein ellipse $B(\Omega,\varrho)$.
\item
\label{cond:MFTExpConvBasket}
The function $(p,z)\mapsto h_p(z)$ is continuous on $\OP\times\Omega$ and there exist functions $H_1:\OP\times\Omega\to\cc$ and $H_2:\Omega\to\cc$ such that for all $(p,z)\in \OP\times\Omega$,
\begin{equation*}
h_{p}(z) = H_1(p,z) H_2(z) 
\end{equation*}
and $H_1(p,z)$ has an extension $H_1:B(\OP,\varrho)\times \Omega\to\cc$ such that, for all fixed $z\in \Omega$ the mapping $p\mapsto H_1(p,z)$ is analytic in the interior of the generalized Bernstein ellipse $B(\OP,\varrho)$.
\end{enumerate}

\subsubsection{Parametric European Options, Generalized Moments and Other Univariate Integrals}
In the generic situation where option prices have to be evaluated for a large set of different parameter constellations, a parametric integral of form \eqref{parametric Integration} for a high dimensional parameter space and a univariate integration domain needs to be computed. This comprises many well-known examples such as prices of European and exotic options and sensitivities of these prices as expressed by the Greeks for different option and model parameters. Also risk measures like VaR and ES and other generalized moments or parametric univariate integrals fall into the scope of this paragraph.

\begin{theorem}
\label{theo:MFTExpConvEU}
Let $\Omega\subset \rr$ and $\OP\subset\rr^{D}$ be compact. Fix some $\eta\in\rr$, some $\varrho>4$ and assume that integrability conditions \emph{(Exp)} and \emph{(Int)} as well as analyticity condition \ref{cond:MFTExpConvEU} are satisfied.
Then for all $p\in\OP$ and $M\in\nn$,
 \begin{align*}
\big\|h_p-I_M(h_p)\big\|_{\infty} &\le CM(\varrho/4)^{-M},
\\
\big|\OI(h_p) - \II_M(h_p)\big|&\le C|\Omega|M(\varrho/4)^{-M},
\end{align*} 
where 
\begin{equation}
C 
= \frac{\varrho}{\varrho-1}\max_{(p,z)\in \OP\times B(\Omega,\varrho)} \big|H_1(p,z)\big| \max_{p\in\OP}|H_2(p)|.
\end{equation}
\end{theorem}
The proof is provided in \cite{GassGlau2015}. 
 In view of a self contained presentation we present the proof in detail in appendix \ref{sec-proofExpo}.

\subsubsection{Basket Options, Multivariate Generalized Moments and Other Multivariate Integrals}
The following result is for example well suited for the error analysis of Magic Point Integration for basket options for a single free parameter. In particular, this is interesting for real-time pricing of basket options with either varying strikes or varying maturities in a fixed calibrated asset model. Moreover, the paragraph applies to the computation of generalized moments such as covariances, and general multivariate integrals with a single varying parameter in the integrand. 

\begin{theorem}
\label{theo:MFTExpConvBasket}
Let $\Omega\subset \rr^d$ and $\OP\subset\rr$ be compact. Fix some $\eta\in\rr^d$, some $\varrho>4$ and assume that integrability conditions \emph{(Exp)} and \emph{(Int)} as well as analyticity condition \ref{cond:MFTExpConvBasket} are satisfied.
Then for all $p\in\OP$ and $M\in\nn$,
 \begin{align*}
\big\|h_p-I_M(h_p)\big\|_{\infty} &\le CM(\varrho/4)^{-M},
\\
\big|\OI(h_p) - \II_M(h_p)\big|&\le C|\Omega|M(\varrho/4)^{-M},
\end{align*} 
where 
\begin{equation}
C 
= \frac{\varrho}{\varrho-1}\max_{(p,z)\in B(\OP,\varrho)\times \Omega} \big|H_1(p,z)\big| \max_{z\in\Omega}|H_2(z)|.
\end{equation}
\end{theorem}
The proof is provided in \cite{GassGlau2015}. Compared to the proof of Theorem \ref{theo:MFTExpConvEU} in appendix \ref{sec-proofExpo}, the only difference is that now the analyticity properties of $H_1$ with respect to the parameters $p$ are exploited to derive an estimate for the best $n$-term approximation of\tild $\OH$.

The implementation of Magic Point Interpolation inevitably involves additional problem simplifications and approximations in order to perform the necessary optimizations. In particular, instead of the whole parameter space a training set is fixed in advance. 
In this context, the results from Theorem\tild \ref{theo:MFTExpConvEU} and\tild \ref{theo:MFTExpConvBasket} are only statements for the training set of functions. Rigorous a priori error bounds for integrals corresponding to parameters outside of the training set can be straightforwardly derived from the a priori error bounds for the Magic Point Interpolation method from \cite{EftangGreplPatera2010}. 

%


\section{Examples and Case Studies}
\label{sec-Ex+Case}

\subsection{Examples of Univariate Payoff Profiles}
\label{sec:MagicFtPayoffs}

Table~\ref{tab:MFTexEuropayoff} presents a selection of payoff profiles $f_K$ for option parameter $K$ as function of the logarithm of the underlying asset. We state the range of possible weight values $\eta$ such that $x\mapsto\ee{\eta x}f_K(x)\in L^1(\IR)$ and the respective generalized Fourier transform exists.
\begin{table}[ht!]
\begin{center}
\begin{tabular}{lcccc}
\toprule\vspace{1ex}
\textbf{Type} & \textbf{Payoff} & \textbf{Weight} & \textbf{Fourier transform}  \\
     &   $f_K(x)$       & $\eta$ & $\widehat{f_K}(z + i\eta)$ & \\[1ex]
     \hline\\
Call         & $(e^x - K)^+$ & $<-1$ & $\frac{K^{iz + 1 + \eta}}{(iz+\eta)(iz + 1 + \eta)}$\\\\
Put  		&  $(K - e^x)^+$ & $>0$ & $\frac{K^{iz + 1 + \eta}}{(iz+\eta)(iz + 1 + \eta)}$ \\\\
Digital & $\mathbbm{1}_{x>\log(K)}$ & $<0$ & $-\frac{K^{iz +  \eta}}{iz + \eta}$ \\
down{\&}out\\[1ex]
Asset-or- & $e^x\mathbbm{1}_{x>\log(K)}$ & $<-1$ & $-\frac{K^{iz + 1 + \eta}}{iz + 1 + \eta}$ \\
nothing\\
down{\&}out\\
\bottomrule\\
\end{tabular}
\caption{Typical payoff profiles for single stock options and the respective generalized Fourier transform.}
\label{tab:MFTexEuropayoff}
\end{center}
\end{table}

Examining the generalized Fourier transforms of the payoff profiles $f_K$ in Table \ref{tab:MFTexEuropayoff}, we realize that all of them admit a factorization in the spirit of condition \ref{cond:MFTExpConvEU} as
\begin{equation}\label{decompo-EU}
\widehat{f_K}(z + i\eta) = K^{iz + c} H_2(z)
\end{equation}
for some $c\in\rr$. 
While all of the payoff profiles $f_K$ of Table \ref{tab:MFTexEuropayoff} either are not differentiable or even discontinuous, the mapping $z\mapsto K^{iz + c}$ is a holomorphic function and thus perfectly fits the requirements of Theorem \ref{theo:MFTExpConvEU}.

\subsection{Example of a Multivariate Payoff Profile}
\label{sec:MagicFtPayoffsBaskets}
The payoff profile of a call option on the minimum of $d$ assets is defined as
	\begin{equation}
		f_K(x) = \left(e^{x_1} \wedge e^{x_2} \wedge \dots \wedge e^{x_d} - K \right)^+,
	\end{equation}
for $x=(x_1,\dots x_d)'\in\mathbb{R}^d$ and strike $K\in\mathbb{R}^+$. With weight value $\eta\in\mathbb{R}^d$, $\eta_j<-1,$ $j=1,\dots d$, the generalized Fourier transform of the multivariate $f_K$ is
	\begin{equation}
	\label{eq:callhatdd}
		 \widehat{f_K}(z+i\eta) = (-1)^d\frac{-K^{1+\sum_{j=1}^d\left(i z_j+\eta_j\right)}}{\prod_{j=1}^d\left(iz_j + \eta_j\right)\left(1+\sum_{j=1}^d\left(iz_j+\eta_j\right)\right)}.
	\end{equation}
A similar decomposition as in \eqref{decompo-EU} in the univariate case can directly be read off. While the mapping $K\mapsto f_K(x)$ displays a kink, the mapping $K\mapsto K^{1+\sum_{j=1}^d\left(i z_j+\eta_j\right)}$ is analytic in the half space $\{K\in\cc\,|\,\Re(K)>0\}$. This perfectly qualifies the call option on the minimum of $d$ assets for the convergence result provided in Theorem \ref{theo:MFTExpConvBasket}.

\subsection{Examples of Asset Models}
\label{sec:MFTModels}

We present a selection of asset models that we use for pricing options in the numerical experiments in section \ref{sec-num} below. The MagicFT algorithm, as we apply it, operates on Fourier integrands that consist of the generalized Fourier transform of the option profile, $\widehat{f_K}$, as well as the Fourier transform of the process that drives the underlying asset at maturity, $\varphi_{T,q}$. Theoretically, Theorem~\ref{theo:MFTExpConvEU} requires the analytic property from the characteristic function $\varphi_{T,q}$ of the model in the sense of condition~\ref{cond:MFTExpConvEU}. Yet, for some models fulfilling this requirement means strongly restricting the parameter space. This would leave us with parameter spaces that are too limited for practical purposes. Empirically, however, we observe that condition\tild \ref{cond:MFTExpConvEU} may be replaced by a much weaker condition while still maintaining exponential convergence. The existence of a shared strip of analyticity $S_R(\eta)$ of width $R\in(0,\infty)^d$ given by
	\begin{equation}
	\label{eq:defMFTstrip}
		S_R(\eta) = \IR^d + i(\eta-R,\eta+R) \subset \IC^d,
	\end{equation}
where all $\xi\mapsto\varphi_{T,q}(\xi)$, $T\in\mathcal{T}$, $q\in\mathcal{Q}$, are analytic on, grants exponential convergence of the algorithm, already. Enforcing such a shared strip means imposing conditions on the model parameter space $\mathcal{Q}$, too. Yet these restrictions turn out to be rather mild compared to the stronger condition~\ref{cond:MFTExpConvEU} of Theorem~\ref{theo:MFTExpConvEU}. 

In the following model presentations we denote by $\widetilde{\mathcal{Q}}$ the parameter space that the model as such is defined on. From this we derive admissible parameter sets $\mathcal{Q}$ such that condition \ref{cond:MFTExpConvEU} is satisfied. If this is not possible, they are chosen to guarantee the existence of a shared strip of analyticity according to\tild \eqref{eq:defMFTstrip}.
Throughout the following model introductions, constant $r>0$ denotes the risk-free interest rate.

\subsubsection{Multivariate \BS Model}
\label{sec:MFTBSd}
The $d$-variate Black-Scholes model is driven by a $d$-variate Brownian motion. The parameter space of the model solely consists of values determining the underlying covariance matrix \text{$\sigma\in\IR^{d\times d}$}, which is symmetric and positive definite. For a concise representation of the parameter space, we define $\widetilde{\mathcal{Q}}$ as
	\begin{equation}
	\label{eq:MFTQSBd}
		\widetilde{\mathcal{Q}} = \{q\in\IR^{d(d+1)/2}\,|\, \det(\sigma(q))>0  \}\subset \IR^{d(d+1)/2}
	\end{equation}
with the function $\sigma:\IR^{d(d+1)/2}\rightarrow \IR^{d\times d}$ defined by
	\begin{equation}
	\label{eq:defMFTsigmafcn}
		\sigma(q)_{ij} = q_{(\max\{i,j\}-1)\max\{i,j\}/2+\min\{i,j\}},\qquad i,j\in\{1,\dots,d\}.
	\end{equation}		
By construction, $\sigma(q)$ is symmetric. The characteristic function of the process\tild $X_T^q$, $T\in\mathcal{T}$, $q\in\widetilde{\mathcal{Q}}$, driving log-returns in the model is then given by
	\begin{equation}
	\label{eq:MFTdCFbs}
		\varphi_{T,q}(z) = \exp\left(T\left( i \langle b ,z\rangle - \frac{1}{2}\langle z , \sigma z\rangle\right)\right),
	\end{equation}
for all $z\in\IR^d$ with drift $b = b(q) \in\IR^d$ adhering to the no-arbitrage condition
	\begin{equation}
	\label{eq:dCFbsdrift}
		 b _i = r-\frac{1}{2}\sigma_{ii},\qquad i\in\{1,\dots,d\}.
	\end{equation}
For each $q\in\widetilde{\mathcal{Q}}$ given by \eqref{eq:MFTQSBd}, the characteristic function of the $d$-variate Black-Scholes model is analytic in $z$ on the whole of $\IC^d$. We thus may choose the parameter set $\mathcal{Q}$ for the MagicFT algorithm according to the following remark.

\begin{remark}[$\mathcal{Q}$ for the multivariate \BS model]
\label{rem:MFTmbsQ}
Let $\underline{\sigma}_i\le\overline{\sigma}_i\in\IR^+$ for all $i\in\{1,\dots,d(d+1)/2\}$. Define
	\begin{equation}
		\mathcal{Q} = \{q\in\IR^{d(d+1)/2}\,|\, \underline{\sigma}_i\le q_i\le \overline{\sigma}_i\text{ such that } \det(\sigma(q))>0 \} 
	\end{equation}		
with the function $\sigma$ given by~\eqref{eq:defMFTsigmafcn}. With the parameter set $\mathcal{Q}$ defined as above and compact $\mathcal{T}\subset \IR^+$, the characteristic function of the univariate \BS model satisfies condition~\ref{cond:MFTExpConvEU} of Theorem~\ref{theo:MFTExpConvEU}.
\end{remark}

\subsubsection{Univariate Merton Jump Diffusion Model}
In the univariate case, the Merton Jump Diffusion model by \cite{merton1976} naturally extends the Black-Scholes model to a jump diffusion setting. The logarithm of the asset price process is composed of a Brownian part with variance $\sigma^2>0$ and a compound Poisson jump part consisting of normally $\mathcal{N}(\alpha,\beta^2)$ distributed jumps arriving at a rate $\lambda>0$. The model parameter space is thus given by
	\begin{equation}
		\widetilde{\mathcal{Q}} = \{(\sigma,\alpha,\beta, \lambda) \in \IR^+\times \IR \times \IR^+_0 \times \IR^+\} \subset \IR^4
	\end{equation}
and the characteristic function of $X^q_{T}$ with $T\in\mathcal{T}$, $q\in\widetilde{\mathcal{Q}}$ computes to
	\begin{equation}
		\label{eq:MFTCFmerton}
		\varphi_{T,q}(z) = \exp\left(T\left(ibz - \frac{\sigma^2}{2}z^2+\lambda\left(\ee{iz\alpha-\frac{\beta^2}{2}z^2}-1\right)\right)\right),
	\end{equation}
for all $z\in\IR$, with no-arbitrage condition
	\begin{equation}
		b = r-\frac{\sigma^2}{2}-\lambda\left(\ee{\alpha+\frac{\beta^2}{2}}-1\right).
	\end{equation}
As in the univariate \BS model, for each $q\in\mathcal{Q}$ and $T>0$, the characteristic function $\varphi_{T,q}$ of the Merton model is holomorphic.

\begin{remark}[$\mathcal{Q}$ for the Merton model]
\label{rem:MFTmertonQ}
Let $\underline{\sigma}\le \overline{\sigma}\in\IR^+$, $\underline{\alpha}\le \overline{\alpha}\in\IR$, $\underline{\beta}\le \overline{\beta}\in\IR_0^+$ and $\underline{\lambda}\le \overline{\lambda}\in\IR^+$. Define
	\begin{equation}
	\begin{split}
		\mathcal{Q} = \{(\sigma,\alpha,\beta, \lambda) \in \IR^4\,|&\ \underline{\sigma}\le \sigma \le \overline{\sigma},\quad \underline{\alpha}\le\alpha\le\overline{\alpha},\\
		&\ \underline{\beta}\le\beta\le\overline{\beta},\quad\underline{\lambda}\le \lambda\le\overline{\lambda} \}.
	\end{split}
	\end{equation}
With the parameter set $\mathcal{Q}$ defined as above and compact $\mathcal{T}\subset \IR^+$, the characteristic function of the Merton model satisfies condition~\ref{cond:MFTExpConvEU} of Theorem~\ref{theo:MFTExpConvEU}.
\end{remark}

\subsubsection{Univariate CGMY Model}
\label{sec:Levy4}
Another well-known L{\'e}vy model that we consider is the univariate CGMY model by Carr, Geman, Madan and Yor (2002)\nocite{CarrGemanMadanYor2002}. This class is also known as Koponen and KoBoL in the literature, see e.g.\ \cite{BoyarchenkoLevendorskiy2002} and as tempered stable  processes. With the model parameter space given by
	\begin{equation}
	\label{eq:MFTcgmyQtilde}
		 \widetilde{\mathcal{Q}} = \{(C,G,M,Y) \in \IR^+ \times \IR_0^+ \times \IR_0^+ \times (1,2)\,|\, (M-1)^Y\!\in \IR\} \subset \IR^4,
	\end{equation}
the associated characteristic function of $X_T^q$ with $T\in\mathcal{T}$, $q\in\widetilde{\mathcal{Q}}$ computes to
	\begin{equation}
	\label{eq:MFTCFcgmy}
	\begin{split}
		\varphi_{T,q}(z) = \exp\big(T&\big(i b  z + C\Gamma(-Y) \\
		&\left[(M-i z )^Y - M^Y + (G+i z )^Y - G^Y\right]\big)\big),
	\end{split}
	\end{equation}
for all $z\in\IR$, where $\Gamma(\cdot)$ denotes the Gamma function. For no-arbitrage pricing we set the drift $ b\in\IR$ to
	\begin{equation}
	\label{eq:CFcgmydrift}
		 b = r - C\Gamma(-Y) \left[(M-1)^Y - M^Y + (G+1)^Y - G^Y\right].
	\end{equation}
The condition $(M-1)^Y\in\IR$ in \eqref{eq:MFTcgmyQtilde} guarantees $b\in\IR$. Contrary to Black-Scholes' and Merton's model, the domain in $\IC$ that the characteristic function of the CGMY model is analytic on does not exist independently of its parametrization. Consequently, Theorem~\ref{theo:MFTExpConvEU} does not apply to pricing in the CGMY model unless the parameter set that the algorithm may choose from is unreasonably restricted. Yet, empirically we maintain exponential convergence in the CGMY model case when $\mathcal{Q}$ and $\eta$ are chosen such that all $\xi\mapsto\varphi_{T,q}(\xi)$, $T\in\mathcal{T}$, $q\in\mathcal{Q}$, share a common strip of analyticity $S_R(\eta)$ as introduced in~\eqref{eq:defMFTstrip} depending on $\eta\in\rr$ and $R>0$, the desired strip width. In the following, we derive conditions which guarantee the existence of such a strip. The result of our analysis will consist in a combined suggestion for the weight value $\eta$ that complies with the restriction posed by the option choice as outlined by Table~\ref{tab:MFTexEuropayoff} and a set of restrictions on the parameter space. These restrictions guarantee a shared strip of analyticity as described above achieving a certain prescribed width $R>0$.

\paragraph{\textbf{Strip of analyticity for CGMY}}
Before we are able to derive conditions on the parameter space that originate a shared strip of analyticity, let us first determine the strip of maximal width $R>0$ that an individually parameterized characteristic function of the CGMY model $\varphi_{T,q}$, $T\in\mathcal{T}$, $q\in\widetilde{\mathcal{Q}}$, is analytic on.

%

\begin{figure}
\centering
\begin{tikzpicture}[
    scale=1,
	axis/.style={very thick, ->, >=stealth'}
    ]    
        \draw[axis] (-0.1,0) --(9.5,0) node(xline)[right] {$\IR$};
        \draw[axis] (0,-2) -- (0, 4) node(yline)[above]{$i\IR$};

	\begin{scope}
        \draw 	(0.1,3) -- (-0.1,3) node[left] {$G$}
            	(0.1,1) -- (-0.1,1) node[left] {$\eta$}
            	(0.1,-1) -- (-0.1,-1) node[left] {$-M$};
    \end{scope}  
  
  	\fill[pattern=north east lines, pattern color=gray] (0,-1) rectangle (8.5,3);
  	
  	\draw[dashed] (0, 1) -- (9,1);
 	\draw (0, 3) -- (9,3) node[right] {$\eta^+$};
 	\draw (0, -1) -- (9,-1) node[right] {$\eta^-$};
 	\draw (2.5,1) node[above, rectangle, fill=white] {$S_R(\eta)$};
 	
 	\draw[<->] (4.5, 1) -- (4.5,3);
	\draw (4.5,2) node[right, rectangle, fill=white] {$R$};
\end{tikzpicture}
\caption{For fixed parametrization $q\in\widetilde{\mathcal{Q}}$, the hatched area visualizes the strip of analyticity of the characteristic function of the CGMY process at $T\in\mathcal{T}$, $X_T^q$. Its bounds are determined by $G\geq 0$ and $M\geq 0$.}
\label{fig:stripanalyticitycgmy}
\end{figure}
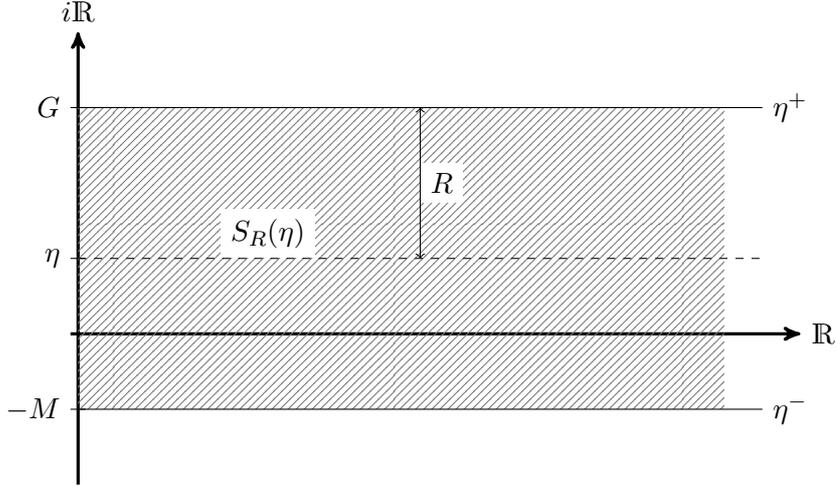

This strip in $\IC$ is derived by analyzing the characteristic function $\varphi_{T,q}$, $T\in\mathcal{T}$, $q\in\widetilde{\mathcal{Q}}$, of the CGMY process on the domain of integration in\tild\eqref{Fourierprice} of Proposition~\ref{prop-FT} for different weight values. Let $\widetilde{\eta}\in\IR$ and consider the characteristic function $\varphi_{T,q}$ on the line
	\begin{equation}
	\label{eq:defMFTcgmyz}
		z_{\widetilde{\eta}}(\xi) = \xi +  i \widetilde{\eta},\qquad \xi\in\IR.
	\end{equation}
The values of $\widetilde{\eta}$ for which $\varphi_{T,q}$ is analytic on the associated line~\eqref{eq:defMFTcgmyz} determine the width of the strip of analyticity of $\varphi_{T,q}$. For these values of $\widetilde{\eta}\in\IR$, both mappings
	\begin{equation*}
	\label{eq:cgmyanalyticxi1}
		\begin{split}
			\xi\mapsto&\ (M-iz_{\widetilde{\eta}}(\xi))^Y,\\
			\xi\mapsto&\ (G+iz_{\widetilde{\eta}}(\xi))^Y
		\end{split}
	\end{equation*}
need to be analytic on $\IR$. By \eqref{eq:defMFTcgmyz}, we have
	\begin{equation*}
		\xi\mapsto(M-iz_{\widetilde{\eta}}(\xi))^Y = (M+\widetilde{\eta}-i\xi)^Y,
	\end{equation*}
and
	\begin{equation*}
		\xi\mapsto(G+iz_{\widetilde{\eta}}(\xi))^Y = (G-\widetilde{\eta}-i\xi)^Y.
	\end{equation*}
For analyticity of these two quantities on $\IR$ we need to ensure that both
	\begin{align}
		M+\widetilde{\eta} >&\ 0, \label{eq:cgmyanalyticxi2}\\
		G-\widetilde{\eta} >&\ 0, \label{eq:cgmyanalyticxi3}
	\end{align}
hold. Inequalities \eqref{eq:cgmyanalyticxi2} and \eqref{eq:cgmyanalyticxi3} yield bounds $\eta^-$, $\eta^+$ given by
	\begin{equation}
	\label{eq:cgmyanalyticxi4}
		\begin{split}
			\eta^+ =&\ G,\\
			\eta^- =&\ -M.	
		\end{split}
	\end{equation}
These two bounds span the strip of analyticity $S_R(\eta)$ for an individually parametrized characteristic function of the CGMY model, wherein $\eta = (\eta^+ + \eta^-)/2 = (G-M)/2$ and diameter $2R = G+M$, as shown in Figure \ref{fig:stripanalyticitycgmy}.


Now we can translate these findings to conditions on the model parameter set to derive a compact set $\mathcal{Q}\subset\widetilde{\mathcal{Q}}$ and a value for $\eta\in\IR$ that ensure a common strip of analyticity $S_R(\eta)$ for all mappings $\xi\mapsto\varphi_{T,q}(\xi)$, $T\in\mathcal{T}$, $q\in\mathcal{Q}$. From our considerations during the derivation above and in particular by \eqref{eq:cgmyanalyticxi4} we conclude that such a $\mathcal{Q}$ and $\eta$ need to satisfy
	\begin{equation}
	\label{eq:cgmyanalyticxi5}
		\max\limits_{(C,G,M,Y)\in\mathcal{Q}} -M < \eta < \min\limits_{(C,G,M,Y)\in\mathcal{Q}} G.
	\end{equation}
We limit the rest of this analysis to the case of a call option where we necessarily have
	\begin{equation}
	\label{eq:cgmyanalyticxi6}
		\eta < -1
	\end{equation}
by Table~\ref{tab:MFTexEuropayoff}. With $G\geq 0$ due to the model parametrization \eqref{eq:MFTcgmyQtilde}, the second inequality in \eqref{eq:cgmyanalyticxi5} trivially holds automatically. Combining \eqref{eq:cgmyanalyticxi5} and \eqref{eq:cgmyanalyticxi6} thus yields condition
	\begin{equation}
	\label{eq:cgmyanalyticxi7}
		\max\limits_{(C,G,M,Y)\in\mathcal{Q}} -M < \eta < -1.
	\end{equation}
A strip width of $R>0$ consequently follows if the final strip condition 
	\begin{equation}
	\label{eq:cgmyanalyticxi8}
		\min\limits_{(C,G,M,Y)\in\mathcal{Q}} M > 1 + 2R
	\end{equation}
is satisfied. In other words, choosing $\mathcal{Q}\subset\widetilde{\mathcal{Q}}$ satisfying condition \eqref{eq:cgmyanalyticxi8} and setting
	\begin{equation}
		\eta = -\min\limits_{(C,G,M,Y)\in\mathcal{Q}}(M+1)/2
	\end{equation}
yields a strip of analyticity $S_R(\eta)$ with diameter $2R$ that all of the mappings $\xi\mapsto\varphi_{T,q}(\xi)$, $T\in\mathcal{T}$, $q\in\mathcal{Q}$, share. We collect and summarize these results in the following remark.

\begin{remark}[$\mathcal{Q}$ for the CGMY model]
\label{rem:MFTcgmyQ}
Let $\underline{C}\le \overline{C}\in\IR^+$, $\underline{G}\le \overline{G}\in\IR_0^+$, $1\le\underline{M}\le \overline{M}\in\IR_0^+$ and $\underline{Y}\le \overline{Y}\in(1,2)$. Let $R>0$ and define
	\begin{equation}
	\label{eq:MFTdefcgmyQ}
	\begin{split}
		\mathcal{Q} = \{(C,G,M,Y) \in \IR^4\,|&\ \underline{C}\le C \le \overline{C},\quad \underline{G}\le G\le\overline{G},\\
		&\ \underline{M}\le M\le\overline{M},\quad\underline{Y}\le Y\le\overline{Y},\\
		&\ (M-1)^Y \in \IR,\\
		&\  M + 2R > 1\}.
	\end{split}
	\end{equation}
All $\varphi_{T,q}$, $T\in\mathcal{T}$, $q\in\mathcal{Q}$, share a common strip of analyticity $S_R(\eta)$ with
	\begin{equation}
		\eta = -\frac{\Big(\min\limits_{(C,G,M,Y)\in\mathcal{Q}} M\Big) +1}{2}.
	\end{equation}	
While the characteristic function of the CGMY model parametrized by\tild $\mathcal{Q}$ of\tild \eqref{eq:MFTdefcgmyQ} in general does not satisfy condition~\ref{cond:MFTExpConvEU} of Theorem~\ref{theo:MFTExpConvEU}, empirically we still observe exponential convergence of the MagicFT algorithm.
\end{remark}


Additionally, to avoid forcing the algorithm to support unrealistic parameter constellations, impose the  following additional plausibility restriction.

\begin{remark}[Plausibility constraint on $\mathcal{Q}$ in the CGMY model]
The implied variance $\sigma_\text{CGMY}^2$ of a CGMY process $(X_t^q)_{t\ge 0}$, $q=(C,G,M,Y)\in\widetilde{\mathcal{Q}}$, at $t=1$ is given by
	\begin{equation*}
		\sigma_\text{CGMY}^2 = C\Gamma(2-Y)\left(\frac{1}{M^{2-Y}} + \frac{1}{G^{2-Y}}\right),
	\end{equation*}
see \cite{CarrGemanMadanYor2002}. For appropriate constants $0<\sigma_- < \sigma_+$ consider imposing the additional condition
	\begin{equation*}
		\sigma_-^2 \le C\Gamma(2-Y)\left(\frac{1}{M^{2-Y}} + \frac{1}{G^{2-Y}}\right) \le \sigma_+^2
	\end{equation*}
for all $(C,G,M,Y)\in\mathcal{Q}$ of Remark~\ref{rem:MFTcgmyQ} thus keeping supported variance levels within reasonable bounds.
\end{remark}

\subsubsection{Univariate Normal Inverse Gaussian Model}
Another L\'evy model we present is the univariate Normal Inverse Gaussian (NIG) model. The parameterization consists of $\delta,\alpha >0$, $\beta\in\IR$, with $\alpha^2 > \beta^2$. The model parameter set\tild $\widetilde{\mathcal{Q}}$ is thus given by
	\begin{equation}
	\label{eq:MFTqtildeNIGuniv}
		\widetilde{\mathcal{Q}} = \big\{(\delta, \alpha, \beta) \in \IR^+\times\IR^+\times\IR\, |\ \alpha^2 > \beta^2,\alpha^2 \geq (\beta + 1)^2\big\} \subset \IR^{3}.
	\end{equation}
The characteristic function of $X_T^q$ for this model is given by
	\begin{equation}
	\label{eq:MFTCFniguniv}
	\begin{split}
		\varphi_{T,q}(z)&=\exp\left(T\left(i b z+\delta\left(\sqrt{\alpha^2 - \beta^2} - \sqrt{\alpha^2 - (\beta+i z)^2}\right)\right)\right)
	\end{split}
	\end{equation}	
for $T\in\mathcal{T}$, $q\in\widetilde{\mathcal{Q}}$, wherein the no-arbitrage condition requires
	\begin{equation}
	\label{eq:CFnigddriftuniv}
		 b = r-\delta\left(\sqrt{\alpha^2 - \beta^2} - \sqrt{\alpha^2 - (\beta + 1)^2}\right).
	\end{equation}
The second condition in \eqref{eq:MFTqtildeNIGuniv}, $\alpha^2 \geq (\beta + 1)^2$, guarantees $b\in\IR$.

As in the CGMY model, the analyticity condition~\ref{cond:MFTExpConvEU} posed by Theorem~\ref{theo:MFTExpConvEU} is not satisfied by all realistic parameter choices $q\in\widetilde{\mathcal{Q}}$. We therefore, analogously to the CGMY case, derive a common strip of analyticity.
\begin{remark}[$\mathcal{Q}$ for the univariate NIG model]
\label{rem:MFTnig1dQ}
Let $\underline{\delta}\le \overline{\delta}\in\IR^+$, $\underline{\alpha}\le \overline{\alpha}\in\IR^+$ and $\underline{\beta} \le \overline{\beta}\in\IR$. Let $R>0$ and define
	\begin{equation}
	\label{eq:MFTq1dNIG}
	\begin{split}
		\mathcal{Q} = \{(\delta, \alpha, \beta) \in\IR^3\,|&\ \underline{\delta}\le\delta\le\overline{\delta},\quad \underline{\alpha}\le\alpha\le\overline{\alpha},\\
		&\ \underline{\beta}\le\beta\le\overline{\beta},\\
			&\ \alpha^2 > \beta^2,\quad\alpha^2 \geq (\beta + 1)^2,\\
			&\ \alpha-\beta > 2R+1,\quad \alpha + \beta > -1\}.
	\end{split}
	\end{equation}
All $\varphi_{T,q}$, $T\in\mathcal{T}$, $q\in\mathcal{Q}$, share a common strip of analyticity $S_R(\eta)$\tild with
	\begin{equation}
		\eta = \frac{\Big(\max\limits_{(\delta, \alpha, \beta)\in\mathcal{Q}}\beta - \alpha\Big) - 1}{2}<-1.
	\end{equation}
While $\mathcal{Q}$ of \eqref{eq:MFTq1dNIG} in general does not satisfy \ref{cond:MFTExpConvEU} of Theorem~\ref{theo:MFTExpConvEU}, empirically we still observe exponential convergence of the MagicFT algorithm.
\end{remark}

\begin{remark}[Plausibility constraint on $\mathcal{Q}$ in the univariate NIG model]
Let $q\in\widetilde{\mathcal{Q}}$ of \eqref{eq:MFTqtildeNIGuniv}. The implied variance $\sigma_\text{NIG}^2$ of a univariate NIG process at $t=1$, $X_1^q$, is given by
	\begin{equation}
		\sigma_\text{NIG}^2(\delta, \alpha, \beta) = \frac{\delta\alpha^2}{\left(\alpha^2-\beta^2\right)^\frac{3}{2}},
	\end{equation}
confer \cite{Prause}. To keep volatilities supported by the MagicFT algorithm within reasonable bounds $0<\sigma_- < \sigma_+$ add the final restriction 
	\begin{equation}
	\label{eq:niganalyticxi11}
		\sigma_-^2 \le \sigma_\text{NIG}^2(q) \le \sigma_+^2,
	\end{equation}
for all $q \in \mathcal{Q}$ of \eqref{eq:MFTq1dNIG}.
\end{remark}

\subsubsection{The univariate Heston Model}

The models introduced above are all L{\'e}vy models. We now introduce the model by \cite{Heston1993} that does not fall into this class but is an affine stochastic volatility model, instead. In the univariate Heston model, the asset price process $(S^q_t)_{t\geq 0}$ follows the stochastic differential equation
	\begin{equation}
	\begin{split}
		\d{S^{q=(v_0,\kappa,\theta,\sigma,\rho)}_t} =&\ rS_t\d{t} + \sqrt{v_t^q}S_t\d{W^1_t},\\
		\d{v^{q=(v_0,\kappa,\theta,\sigma,\rho)}_t} =&\ \kappa(\theta-v_t)\d{t} + \sigma\sqrt{v^q_t}\d{W^2_t},
	\end{split}
	\end{equation}
with the two Brownian motions $W^1$, $W^2$ correlated by $\rho\in[-1,1]$ and with $q\in\widetilde{\mathcal{Q}}$ defined by	
	\begin{equation}
	\label{eq:defMFThestonQtilde}
	\begin{split}
		\widetilde{\mathcal{Q}} = \big\{(v_0, \kappa, \theta, \sigma, \rho)\in \IR^+\!\times \IR^+\! \times \IR^+\! \times \IR^+\! \times [-1,1], \sigma^2 \le 2\kappa\theta\big\}.
	\end{split}
	\end{equation}	
The Feller condition 
	\begin{equation*}
		\sigma^2 \le 2\kappa\theta
	\end{equation*}
in $\widetilde{\mathcal{Q}}$ of \eqref{eq:defMFThestonQtilde} ensures an almost surely non-negative volatility process $(v_t)_{t\geq 0}$. With $T\in\mathcal{T}$, $q\in\widetilde{\mathcal{Q}}$, the characteristic function $\varphi_{T,q}$ of the log-asset price process $(\log(S_t/S_0))_{t\geq 0}$ at $T$ is given by
\begin{equation}
	\label{eq:HestonChar}
	\begin{split}
		 \varphi_{T,q}(z) =& \exp\left(T\,irz\right)\exp\bigg(\frac{v_0}{\sigma^2}\frac{(a-c)(1-\exp(-cT))}{1-g\exp(-cT))} \\
		&\qquad +\frac{\kappa\theta}{\sigma^2}\left[(a-c)T - 2\log\left(\frac{1-g\exp(-cT)}{1-g}\right)\right]\bigg),
	\end{split}
	\end{equation}
for all $z\in\IR$, with supporting functions defined by
	\begin{equation*}
	\begin{split}		
		a = a(z) =&\ \kappa-i\rho\sigma z,\\
		c = c(z) =&\ \sqrt{a(z)^2-\sigma^2(-zi-z^2)},\\
		g = g(z) =&\ \frac{a(z)-c(z)}{a(z)+c(z)},
	\end{split}
	\end{equation*}
confer \cite{Schoutens2004}. We simply choose $\mathcal{Q}\subset\widetilde{\mathcal{Q}}$ to be a bounded subset of the parameter space.
\begin{remark}[$\mathcal{Q}$ for the univariate Heston model]
\label{rem:MFThestonQ}
Choose bounds for the initial value of the volatility process, $0<\underline{v_0} \le \overline{v_0}$, for its speed of mean reversion, $0<\underline{\kappa}\le\overline{\kappa}$, the long-term volatility mean, $0<\underline{\theta}\le\overline{\theta}$, and the volatility of the volatility process itself, $0<\underline{\sigma}\le\overline{\sigma}$, and a domain for the correlation parameter, $-1\le\underline{\rho}\le\overline{\rho}\le 1$. Define
	\begin{equation}
	\label{eq:defMFThestonQ}
	\begin{split}
		\mathcal{Q} = \big\{(v_0, \kappa, \theta, \sigma, \rho)\ |\ &\underline{v_0} \le v_0 \le \overline{v_0},\ \underline{\kappa}\le\kappa\le\overline{\kappa},\\
		& \underline{\theta}\le\theta\le\overline{\theta},\ \underline{\sigma}\le\sigma\le\overline{\sigma}, \underline{\rho}\le\rho\le\overline{\rho},\\
		& \sigma^2 \le 2\kappa\theta\big\}.
	\end{split}
	\end{equation}
Despite the fact that $\mathcal{Q}$ defined above in general might not satisfy condition~\ref{cond:MFTExpConvEU} of Theorem~\ref{theo:MFTExpConvEU}, we still observe exponential convergence of the MagicFT algorithm.
\end{remark}

For an analysis of the strip of analyticity in the Heston model, see \cite{Levendorskiy2012}.
\section{Numerical Experiments}\label{sec-num}

In the previous sections we introduced the MagicFT algorithm for option pricing and presented several asset models and option types. We also proved theoretical claims for option pricing with the MagicFT algorithm. In this section we numerically validate these theoretical claims and provide empirical indication that the scope of the algorithm extends to a much wider class of pricing applications than suggested by the theorems earlier.

\subsection{Implementation}
\label{sec:MagicFtImplementation}

The implementation of the algorithm in Matlab introduces some simplifications as suggested by e.g. Remark~3 in \cite{MadayNguyenPateraPau2009}. A theoretical argumentation for the discretization approach described in the following can be found in \cite{EftangGreplPatera2010} and  \cite{Maday14GEIM}. The continuous parameter space $\mathcal{P}$ is thus replaced by a discrete parameter cloud randomly sampled. Each magic parameter that the algorithm selects is a member of this discrete set. Consequently, the set $\mathcal{U}$ that the algorithm is trained on is replaced by a discrete set, as well. 
Additionally, we take $\Omega$ to be a discrete set with a finite number of points in each spacial dimension. 
Each function $u\in\mathcal{U}$ is then represented by its evaluation on this discrete $\Omega$ and is thus replaced by a finite-dimensional vector, numerically. 
The optimization steps from \eqref{def_u1}--\eqref{defxi_M} thus reduce to a search on finite sets.
When all $h_{p^\ast_m}\in \OU$ for $m=1,\ldots,M$ are identified, they are integrated using Matlab's \texttt{quadgk} routine (with an absolute tolerance requirement of $10^{-14}$, a relative tolerance requirement of $10^{-12}$, a maximum number of intervals of $200,000$) and linearly assembled to derive the quantities $\int_\Omega \theta^M_m(z)\dd z$ for $m=1,\ldots,M$.

\begin{table}[htb!]
\begin{center}
\begin{tabular}{@{}lllllll@{}}
\toprule
\textbf{Model}  & \phantom{a} & \multicolumn{2}{c}{\textbf{fixed parameters}}                  & \phantom{a} & \multicolumn{2}{c}{\textbf{free parameters}}                \\ 
\midrule
BS   &                 & $K=1$                     &             &           & $S_0/K\in[0.5,\ 2]$,    &  $T\in[0.1,\ 1.5]$,  \\
	& 		 & 				 &		&     	& 	 $\sigma\in[0.1,\ 0.9]$ \\
\midrule
Merton    	&                 & $K=1$                &              &          &  $S_0/K\in [0.5,\ 2]$,    &  $T\in [0.1,\ 1.5]$,      \\
       		&                 &                           &              &          &   $\sigma\in [0.1,\ 0.7]$, & $\alpha\in [-1.5,\ -0.1]$,    \\
       		&                 &                           &              &          &   $\beta\in [0.1,\ 1]$,         & $\lambda\in [10^{-5},\ 1]$  \\
\midrule
NIG    	&                 & $K=1$                &             &         &   $S_0/K\in[0.5,\ 2]$,    &  $T\in[0.1,\ 1.5]$,                          \\
       		&                 &                           &             &          &   $\alpha\in[10^{-5},\ 3]$, & $\beta\in[-3,\ 3]$,                         \\
       		&                 &                           &             &          &   $\delta\in[0.2,\ 1]$    &   \\
\midrule
CGMY   	&                 & $K=1$,               &   $Y=1.1$    &          & $S_0/K\in[0.5,\ 2]$,    &  $T\in[0.1,\ 1.5]$,    \\
       		&                 &                           &                    &          &  $C\in[10^{-5},\ 1]$, &   $G\in[0,\ 25]$,                       \\
       		&                 &                           &                    &          &    $M\in[0,\ 30]$    &                           \\
\midrule
Heston 	&                 & $K=1$,               	& $\kappa=2$,           &          &  $S_0/K\in[0.5,\ 2]$,    &  $T\in[0.1,\ 1.5]$,  \\
       		&                 &  $\sigma=0.15$   &       			&          &    $v_0\in[0.2^2,\ 0.3^2]$,  &    $\theta\in[0.15^2,\ 0.35^2]$,                      \\
       		&                 &                           &             			&          &      $\rho\in[-1,1]$                     &                           \\
\bottomrule
\end{tabular}
\caption{In the numerical experiments, we price European call options as an example. Various models have been selected. In the implementation, the Fourier integrands that the algorithm constructs the basis functions $q_m$ with are parametrized according to the intervals above. For each model investigated, $\mathcal{U}$ consists of a pool of $|\mathcal{U}|=4000$ Fourier integrands.}
\label{tab:MagicFTEuroCallparam}
\end{center}
\end{table}
\subsection{Empirical Convergence}
\label{sec:MagicFtEmpiricalConvergence}

We study the empirical convergence of our implementation of the MagicFT pricing algorithm. A plain vanilla European call option on one asset serves as an example. We investigate the convergence in several models. For each model we set up a pool $\mathcal{U}$ of parametrized Fourier integrands that the algorithm picks from. For each model, the discrete parameter pool is chosen as a uniform sample of magnitude $|\mathcal{P}|=4000$ from the free parameter ranges enlisted in Table~\ref{tab:MagicFTEuroCallparam}.  

It is interesting to note that, not necessarily all model parameters have to be considered in the parametric option pricing. For example the parameter $Y$ in the CGMY model reflects the degree of roughness of the paths of the process and can in principle be estimated from the historical stock price data. In contrast to other model parameters, such as those determining variance and skewness, the path behaviour is generally consider be stable over time. We can therefore fix the parameter $Y=1.1$ in the CGMY model. Additionally, for the NIG and CGMY model, a shared strip of analyticity of width $R=1/2$ is enforced such that for all investigated models, the dampening factor $\eta$ could be set to $\eta=-1.5$. Furthermore, all model restrictions stated in section \ref{sec:MFTModels} are respected. Also, implied variances are kept in the interval $[0.01^2, 0.8^2]$. Each Fourier integrand is evaluated on a discrete $\Omega \subset [0,65]$ with $|\Omega|=1714$.
\begin{figure}[htb!]
\centering
\makebox[0pt]{\includegraphics[scale=1]{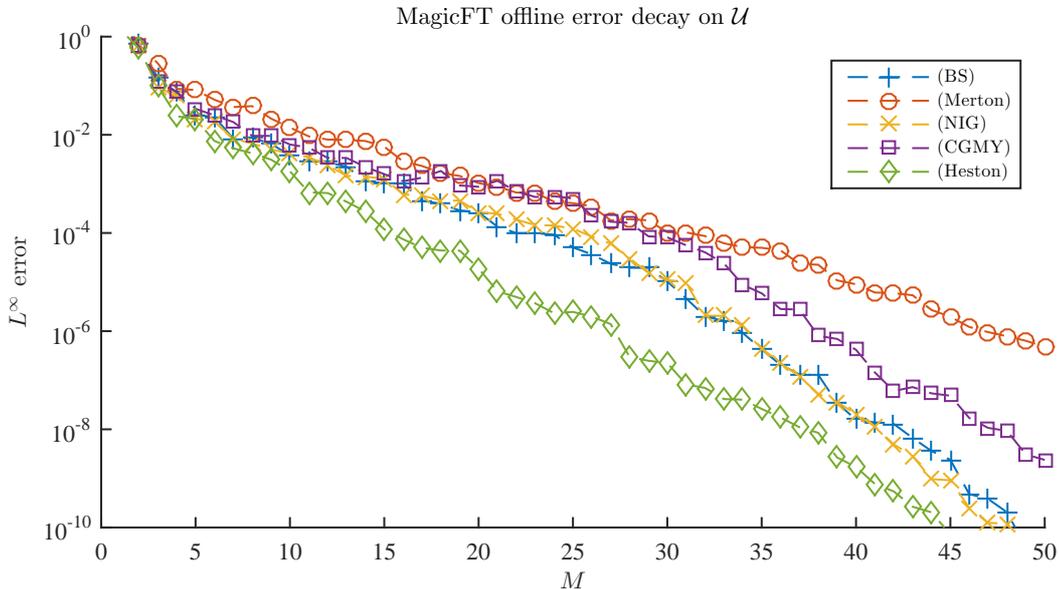}}
\caption{A study of the empirical order of convergence of the error in step~\eqref{defxi_M} during the offline phase of the MagicFT algorithm. Five different models and European call options are considered. Both the models and the option are parametrized according to Table~\ref{tab:MagicFTEuroCallparam}. The convergence result is theoretically backed by Theorem~\ref{theo:MFTExpConvEU} for the \BS and the Merton model. A shared strip of analyticity of the respective Fourier integrands of width $R=1/2$ has been enforced for the NIG and CGMY model. 
}
\label{fig:MagicFTerrordecay}
\end{figure}

Figure~\ref{fig:MagicFTerrordecay} shows the empirically observed error decay during the offline phase of the algorithm for all five considered models in the number of basis functions\tild $M$. For each model, the quantity $\underset{z\in \Omega}{\max}\big|u_M(z) - I_{M-1}(u_M)(z)\big|$ is shown for increasing values of $M$. The algorithm has been instructed to construct basis functions $q_m$ until an error threshold of $10^{-10}$ has been reached in step~\eqref{defu_M} or until $M$ has reached the value $50$. This offline phase based on the Matlab implementation described in the previous Section~\ref{sec:MagicFtImplementation} takes less than 1 minute of time on a standard laptop computer for each model.

We observe exponential error decay in all considered models. Recall that Theorem~\ref{theo:MFTExpConvEU} predicts this behavior only for the \BS and the Merton model where analyticity of the associated Fourier integrands is parameter independent. For the other two L{\'e}vy models, however, the existence of a shared strip of analyticity results in exponential error decay, as well. In case of the Heston model, the issue of analyticity of the Fourier integrands in $\mathcal{U}$ has not been investigated here. Still, we observe exponential error decay too. The empirical results depicted in Figure~\ref{fig:MagicFTerrordecay} thus indicate that it might be promising to investigate a theoretical result providing exponential error decay  beyond the scope of Theorem~\ref{theo:MFTExpConvEU}.

\subsection{Out of sample pricing study}\label{sec-out-of-sample}

In the previous paragraph we studied empirical convergence during the offline phase of the algorithm. More precisely, we investigated for several models how accurately all Fourier integrands in the given pool $\mathcal{U}$ could be approximated on their integration interval $\Omega$ by the $M$ selected integrands or rather by the basis functions $q_m$, $m=1,\dots,M$, constructed thereof. Now we analyze, how the observed accuracy on the level of in sample integrands translates to the accuracy in an out of sample call option pricing exercise.

\begin{figure}[htb!]
\centering
\makebox[0pt]{\includegraphics[scale=1]{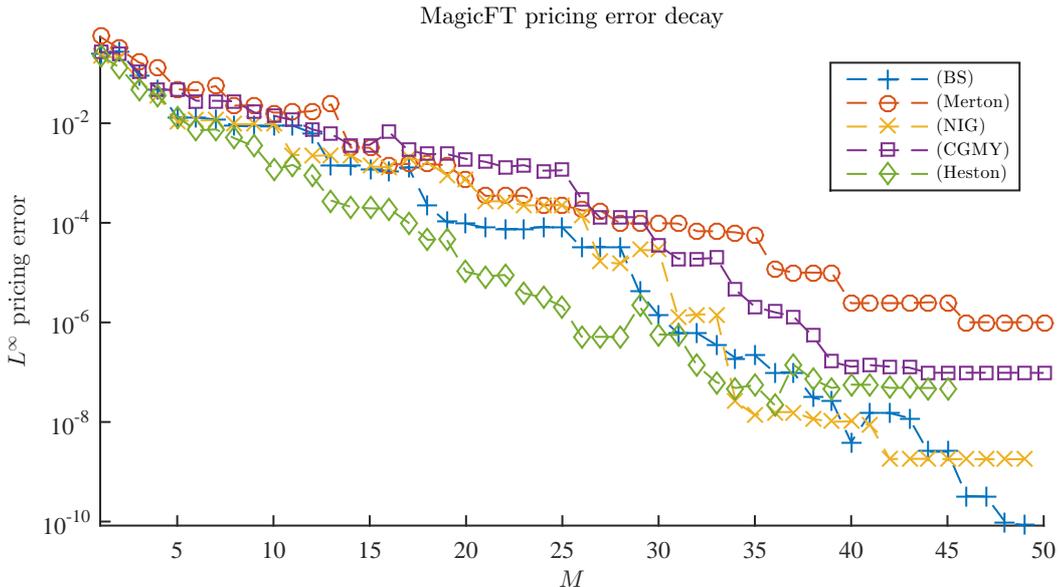}}
\caption{Pricing error decay study on 1000 out of sample parameter constellations for different models. In each model, for increasing values of $M$, the $L^\infty$ error over the randomly drawn parameter sets is evaluated. The parameter sets have been drawn from the intervals given by Table~\ref{tab:MagicFTEuroCallparam}.}
\label{fig:MagicFTerrordecayPrices}
\end{figure}

\begin{figure}[htb!]
\begin{center}
\makebox[0pt]{\includegraphics[scale=0.935]{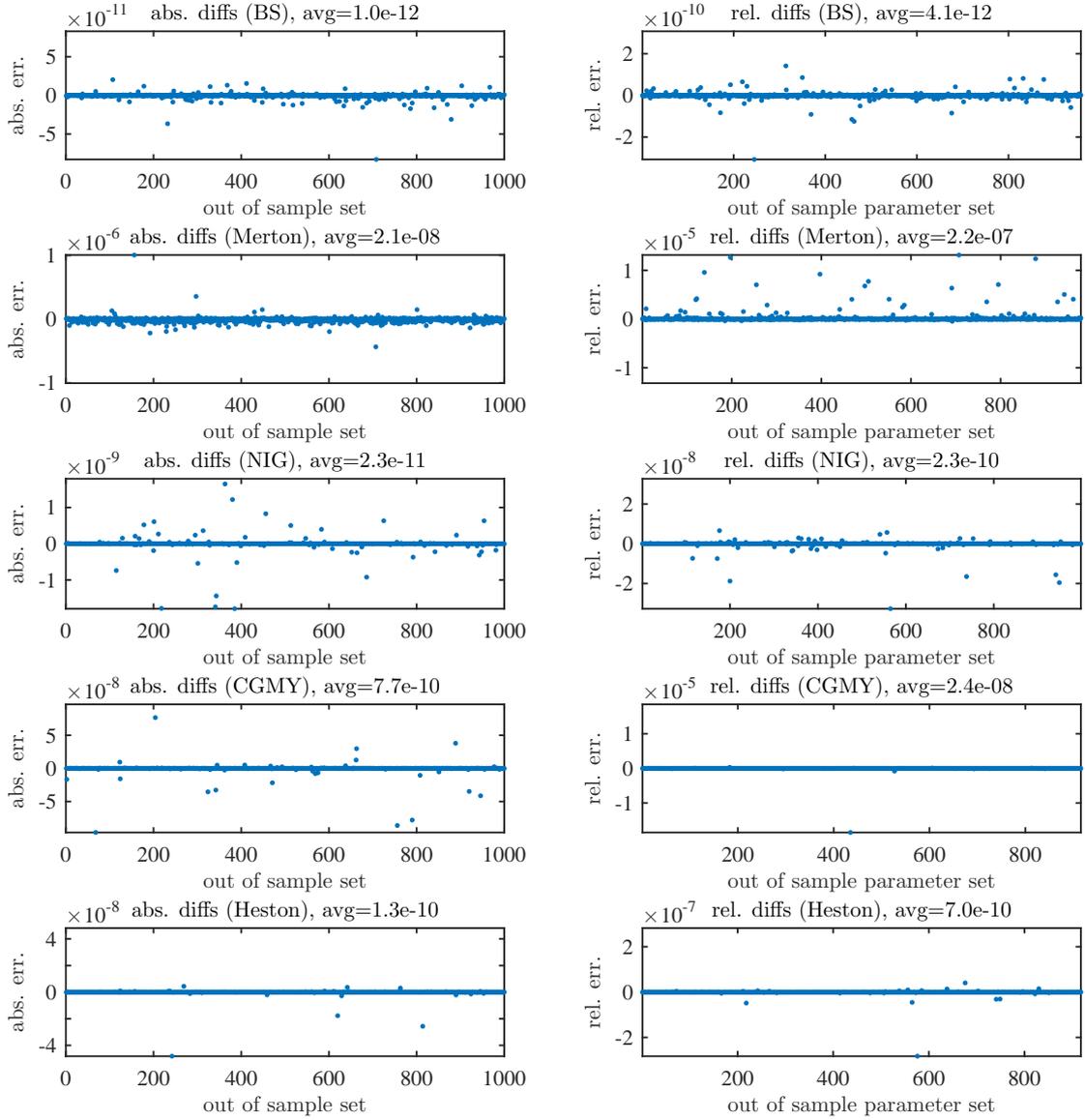}}
\end{center}
\caption{Results of the out of sample pricing exercise. For each of the five considered models, 1000 parameter sets have been drawn from the intervals given by Table~\ref{tab:MagicFTEuroCallparam}. For each set, the Fourier price as well as the MagicFT price have been calculated. On the left column, all absolute errors are depicted. On the right, the relative errors are shown.}
\label{fig:MagicFToutofsampleerrs}
\end{figure}

To this extent we randomly draw 1000 parameter constellations for each model according to the same rules as in the offline phase. For each such sample we compute the respective Fourier price by numerical integration on $[0,65]$ thus containing the discrete $\Omega$ that the MagicFT algorithm has been trained on. We integrate using Matlab's \texttt{quadgk} with absolute tolerance of $10^{-12}$ and $200,000$ integration intervals. Additionally, in each model we approximate all prices associated with the randomly drawn parameters for increasing values of\tild $M$, evaluate the $L^\infty$ error and study its decay in $M$ as depicted in Figure~\ref{fig:MagicFTerrordecayPrices}.

We observe exponential rates for all considered models. Curiously, the error decay attains plateau-like shapes, especially for higher values of $M$. We explain this decay structure by assuming that each plateau is associated with a certain single parameter realization from the random sample that dominates the $L^\infty$ error until a magic parameter close to it or rather the respective basis function contributes to the approximation of the belonging price. Due to such outliers, the order in which the offline phase errors were decaying in Figure~\ref{fig:MagicFTerrordecay} has changed.

In Figure~\ref{fig:MagicFToutofsampleerrs}, we depict evaluations of the absolute as well as the relative pricing errors for all out of sample parameter sets, individually. Here, relative errors have been computed only for prices larger than $10^{-3}$ to exclude numerical noise. In each model, $M$ is set to its final value assigned during the respective model's offline phase and can be read off from Figure~\ref{fig:MagicFTerrordecay}.


Pricing accuracy in this out of sample pricing exercise reaches very satisfactory levels albeit the achieved accuracies vary between the considered models. For all models, average absolute pricing accuracy reaches levels between $\text{avg}_\text{min} \approx 10^{-12}$ in the \BS model and $\text{avg}_\text{max} \approx 10^{-8}$ for the Merton model. Average relative pricing accuracy ranges between $10^{-12}$ and $10^{-7}$. We observe individual outliers for all models.
Even occasional mispricing, however, stays within practically acceptable bounds smaller than $10^{-5}$. The ten worst absolute errors in each model are further addressed in the next section.

\subsection{Individual Case Studies}

We take a closer look into the numerical results for each model individually.  For this we are interested in the distribution of magic parameters in each model.

\paragraph{\textbf{Black-Scholes}}

During the offline phase of the algorithm for the Black-Scholes model only the option strike $K$ has been fixed, $K=1$. The model parameter\tild $\sigma$ as well as the two other parameters $S_0/K$ and maturity $T$ were allowed to vary within the bounds assigned by Table~\ref{tab:MagicFTEuroCallparam}. In the \BS case, the individual parameter intervals tensorize meaning that any combination of parameter values respecting the individual bounds can be picked by the algorithm.
\begin{figure}[htb!]
\centering
\subfigure
{\includegraphics[width=0.49\textwidth]{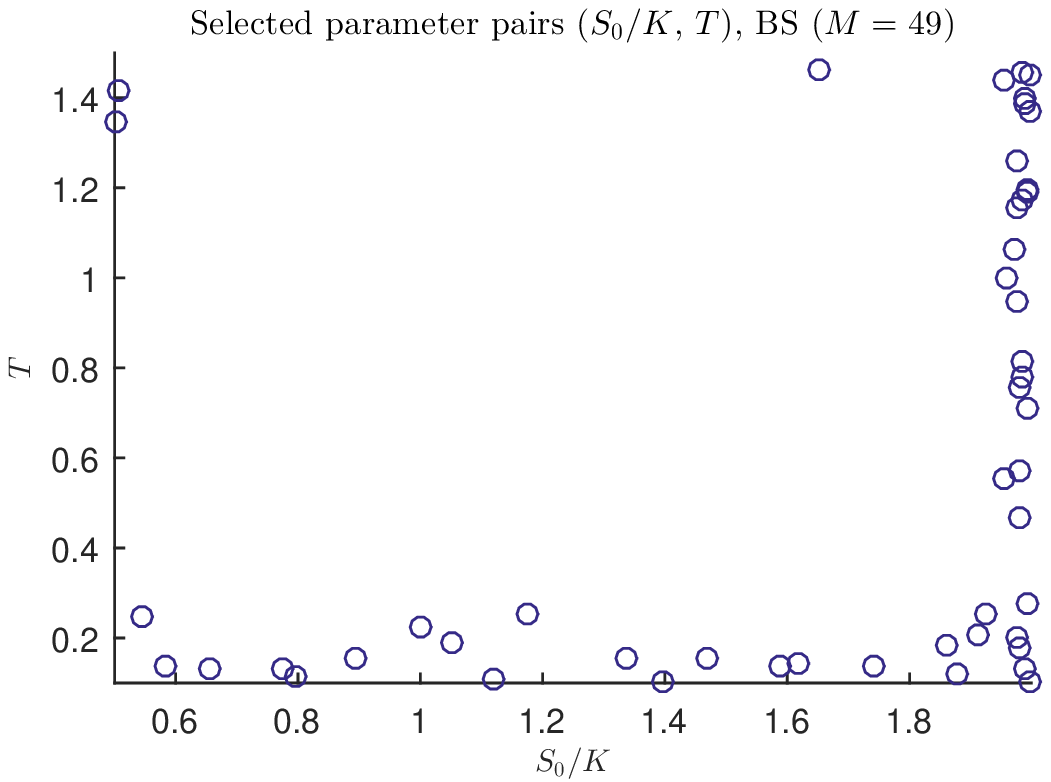}}
\subfigure
{\includegraphics[width=0.49\textwidth]{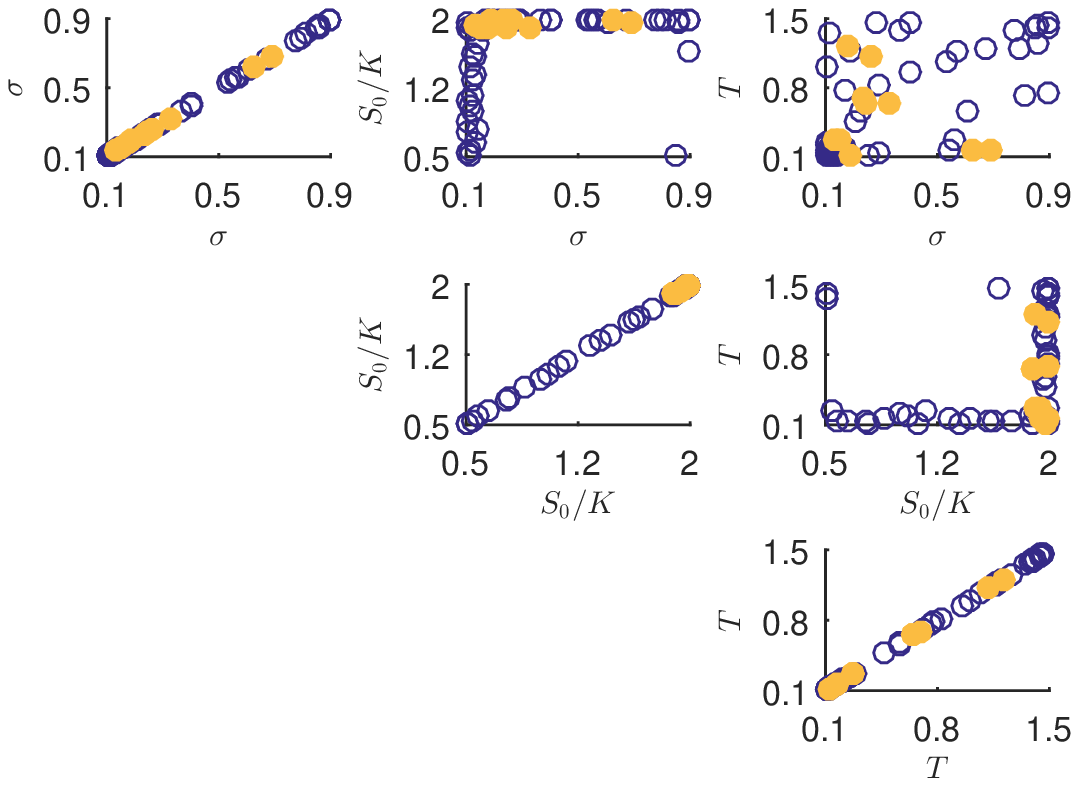}}
\caption{Left: Parameter pairs $(S_0/K,T)$ selected by the MagicFT algorithm in the offline phase of the \BS model. Right: All magic parameters selected during the offline phase of the algorithm for the \BS model (empty blue circles). The filled orange circles denote the ten parameter constellations that resulted in the maximal absolute pricing errors during the out of sample pricing exercise.}
\label{fig:MagicFTBSparcloudSKT_MagicFTBSCloudWithMaxErr}
\end{figure}

As the left part in Figure~\ref{fig:MagicFTBSparcloudSKT_MagicFTBSCloudWithMaxErr} demonstrates for the magic parameter choices for $S_0/K$ and $T$, however, rather extreme constellations have been selected. The right part of Figure~\ref{fig:MagicFTBSparcloudSKT_MagicFTBSCloudWithMaxErr} provides a complete overview over all parameter combinations selected in the offline phase of the algorithm for the Black-Scholes model. With the exception of $T$ and $\sigma$ combinations, rather extreme parameter pairs have been selected. This special behavior is not surprising, since $T$ and $\sigma$ always appear together as a product in the Fourier integrands of the \BS model, compare the definition of the characteristic function in the \BS model in~\eqref{eq:MFTdCFbs}. The even distribution of the $(T,\sigma)$ parameter pairs thus reflects the even distribution of all individual parameters over their domain, observable on the elements on the main diagonal of the figure.

\subsection{Comparison with the Cosine method}

A wide range of different methods for the efficient evaluation of Fourier integrals for option pricing have been successfully applied. One of these methods is the popular cosine method of \cite{FangOosterlee2008}.
We use the cosine method as relevant benchmark to our MagicFT method, also since 
an implementation for the Black-Scholes, Heston and Merton model (among others) by the original authors is
publicly available in the BENCHOP project, see \cite{Sydow2015}. 
Despite their similarities as Fourier pricing routines, both methods differ conceptually since MagicFT is an offline-online scheme.


In order to compare the MagicFT and the cosine method for the Black-Scholes, Heston and Merton model, we use for each model the parameter sample set from Section \ref{sec-out-of-sample}.
The accuracies of both methods will be measured against the Fourier integral 
\begin{align}
\frac{1}{2\pi} \int_{\Omega + i\eta }  \widehat{f_K}(-z) \varphi_{T,q}(z) \dd z\label{eq:fourier_integral_bounded}
\end{align}
with $\Omega=[0,65]$ as benchmark.

To allow the comparison of both methods, we will only consider parameter constellations for which this benchmark does not exhibit truncation errors above a threshold of $\varepsilon^\text{params} > 10^{-8}$. Therefore 11 of the 1000 parameter tuples have been omitted according to the criterion
\begin{align*}
\frac{1}{2\pi} \bigg|\int_{\Omega^c + i\eta }  \widehat{f_K}(-z) \varphi_{T,q}(z) \dd z\bigg| >\varepsilon^\text{params}.
\end{align*}

The cosine method allows to set a specific integration range via the parameter $L$ in equation (49) of \cite{FangOosterlee2008}. As mentioned in \cite{InnocentisLevendorskiy2014} this parameter needs to be carefully selected to guarantee convergence of a whole range of pricing parameters. In the preparation of our numerical studies, we have identified that the parameter $L=14$, $L=18$ and $L=3.1$ for the models Black-Scholes, Heston and Merton, respectively, lead to the best possible convergence results.

\begin{figure}[htb!]
\centering
\subfigure
{\includegraphics[width=0.495\textwidth]{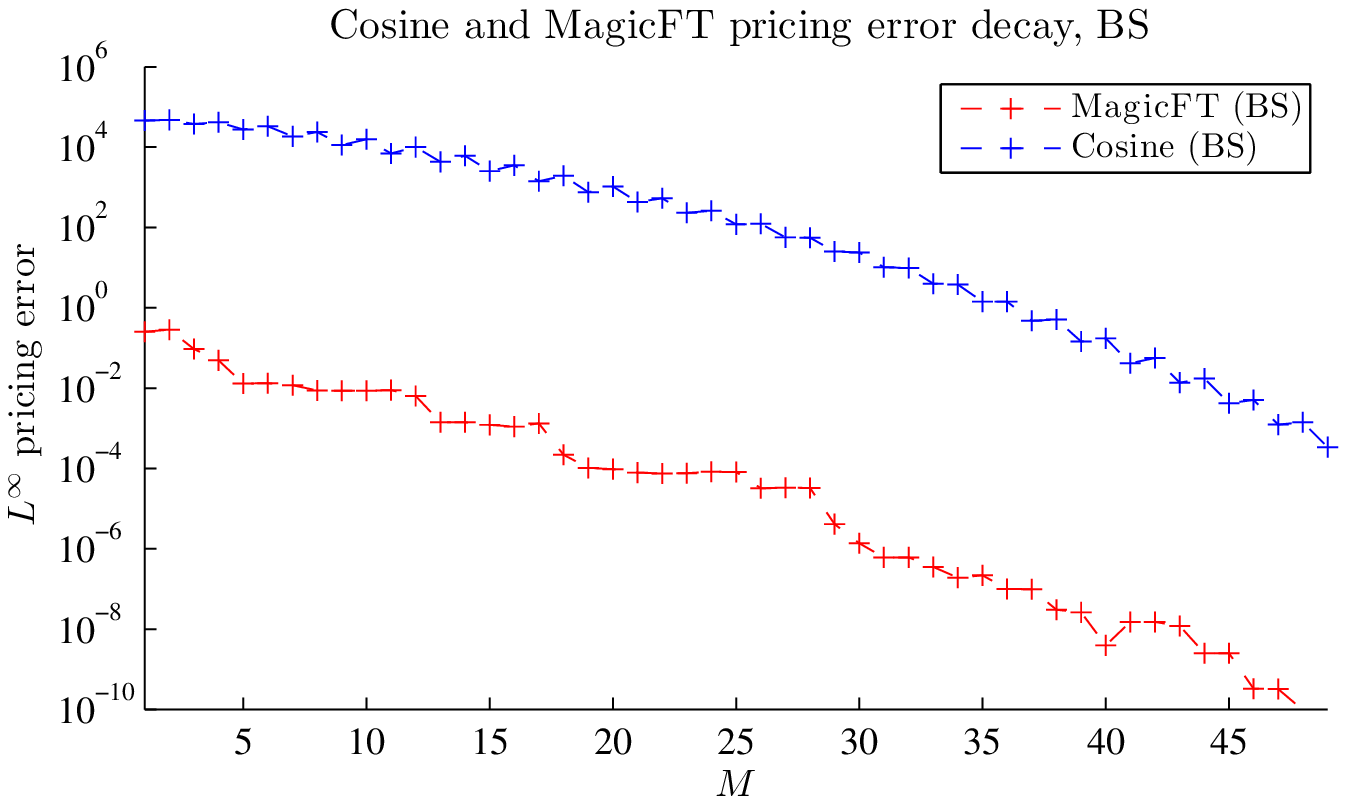}}
\subfigure
{\includegraphics[width=0.495\textwidth]{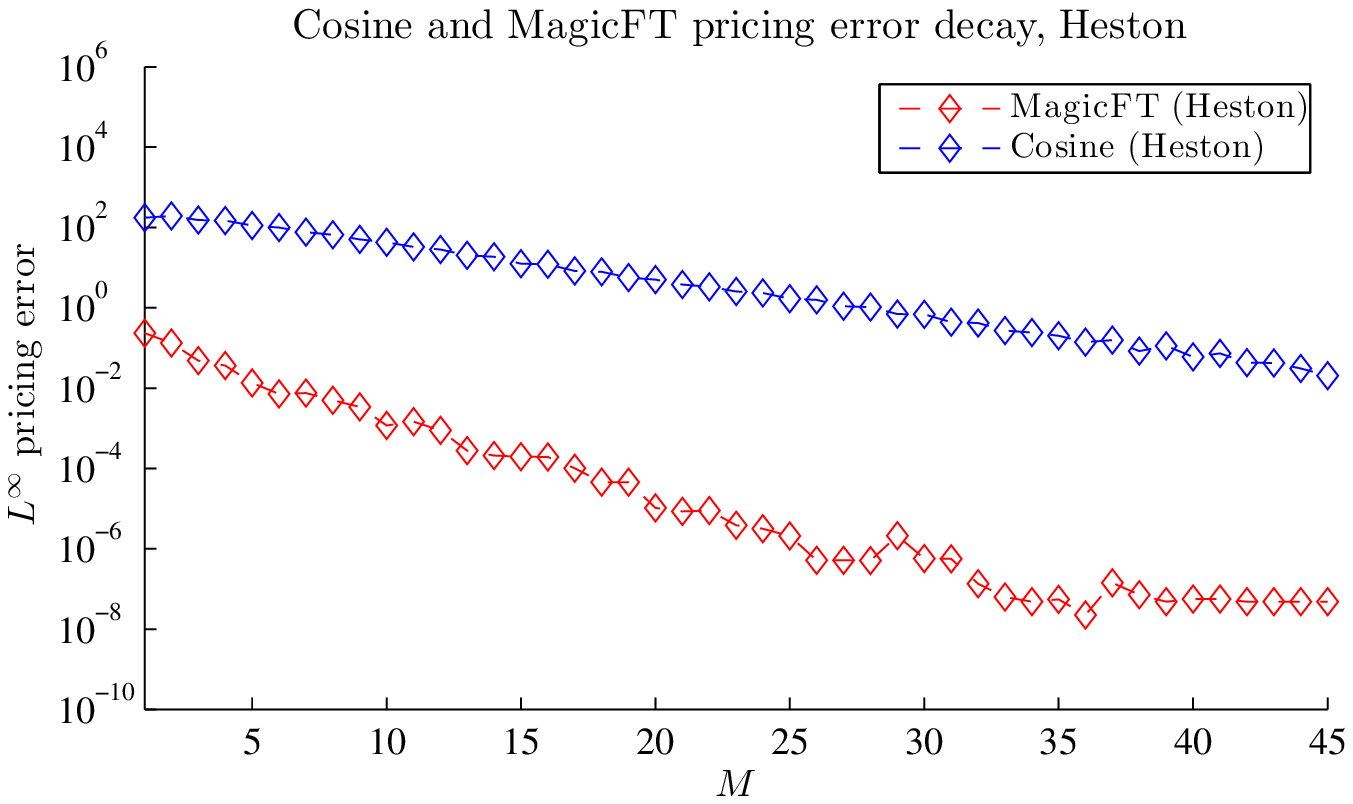}}
\subfigure
{\includegraphics[width=0.495\textwidth]{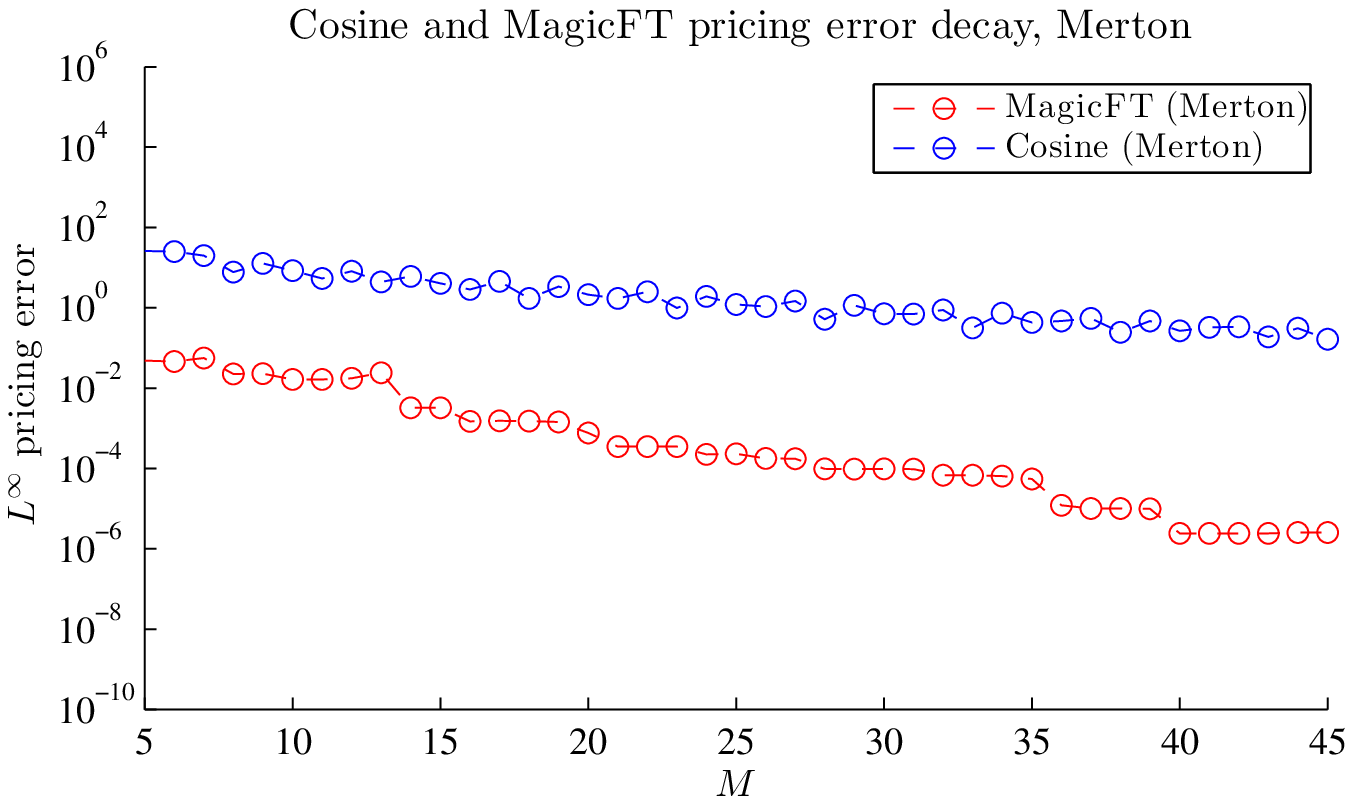}}
\caption{Efficiency study of the MagicFT and cosine method for the Black-Scholes, Heston and Merton models. The plots show the $L^\infty$ error across 1000 random parameter constellations.}
\label{fig:COS_MagicFT_compare}
\end{figure}

The accuracy of both numerical methods for varying numbers of nodes $M=1$ to $M=50$ is shown in Figure~\ref{fig:COS_MagicFT_compare}. Instead of comparing milliseconds in CPU time we use the number of summands $M$ of both methods as a measure for the computational complexity of both pricing routines. We deem this approach justified by the fact that pricing in both methods consists of assembling sums of known coefficients multiplied essentially by the characteristic function of the underlying model, which is available in closed form for all examples considered. Thus Figure~\ref{fig:COS_MagicFT_compare} can be directly used to infer the efficiency of the numerical methods. The plots show that the errors of the MagicFT method is significantly below the error of the cosine method from $M=15$ onwards. In the Black-Scholes case, both methods show a similar rate of convergence, while MagicFT is more accurate in absolute values. In the Heston model, both methods show an exponential rate of convergence. Here, MagicFT exhibits a higher rate and is more accurate.  For the Merton model, we observe that the cosine method is not accurate for the whole range of parameters. To give a quantitative example: for $M=50$, the error of MagicFT reaches levels of the order of $10^{-6}$, whereas the cosine method still shows errors of $3.8\cdot 10^{-4}$ for the Black-Scholes, $1.1\cdot 10^{-2}$ for the Heston and $0.13$ for the Merton model.

As mentioned before, the literature on Fourier-based pricing offers many other reliable and efficient approaches to evaluate Fourier integrals. In particular, \cite{Levendorskii2016} proposes to choose an appropriate deformation of the contours of integration prior to the discretization. This approach leads to very accurate results already for few discretization points 
and thus is especially attractive.  
 It is interesting to consider a combination of the MagicFT method with the the approach of \cite{Levendorskii2016}. Both methods thereby would benefit mutually and we expect further gains in efficiency. To be more precise, we suggest to first  choose the contour of integration optimally in regard to analyticity. Then, as a second step the MagicFT algorithm can be applied for the resulting parametric integrals. This would allow MagicFT to benefit from the improved region of analyticity of the integrands.


\section{Outlook}

The results of the experiments for univariate Fourier integrals indicate that extensions of MagicFT to multivariate option pricing is promising. Firstly, the univariate method benefits from the offline phase when compared to the cosine method. Secondly, the architecture of this offline-online method guarantees a rate of convergence that does not intrinsically suffer from the curse of dimensionality since it directly relates to the best $n$-term approximation. The core of the offline phase is an optimization procedure, that in contrast will directly be affected by the curse of dimensionality. It is, however, crucial to notice that the offline phase only needs to be performed once for a whole model class and option type. Therefore the run-time of the offline phase can be seen as part of the implementation phase of the pricing routine. For all of our univariate examples the offline phase has required only one minute on a standard laptop. For moderate dimensions we therefore expect that a direct extension of the algorithm will still lead to a practically useful method. 

Our implementation of the offline phase of the MagicFT algorithm follows the standard procedure in the literature. Here a global optimization routine that requires the evaluation of the integrands for a large number of samples in the parameter space is used to find the magic points and basis functions. This optimization algorithm exhibits reasonable run-times in our numerical examples. 
For applications in higher dimensions, refinements of this implementation of the offline phase can be beneficial and are currently being investigated in the empirical integration literature, see \cite{Maday14GEIM}.

\section{Conclusion}

We have introduced the MagicFT algorithm for parametric option pricing (POP). Analyticity conditions theoretically guarantee an exponential rate of convergence of the method in the number of magic points. The numerical experiments confirm these findings and suggest an exponential rate of convergence even for models and options beyond the scope of our theoretical results. This gives rise to the hope that further valuable theoretical results can be established.

Thanks to its architecture, the method is highly efficient for a pre-specified \textit{range of parameter constellations} of interest. 
In contrast to other interpolation methods, there are no generic geometric constraints for the  choice of the parameter space. 
We have compared experimentally the performance of the MagicFT method to the cosine method for a whole range of parameters. This comparison indicates that the MagicFT method is beneficial when the efficiency for a whole range of parameters is crucial.


\appendix
\section{Properties of Magic Point Interpolation}\label{sec-insight-magic}
For the reader's convenience we state useful features of the algorithm and give a detailed proof of the convergence result Proposition \ref{prop-conv}, which basically coincides with Theorem 2.4 of \cite{MadayNguyenPateraPau2009}.

\subsection{General features}
\label{sec:MFTGeneralFeatures}


The Magic Point Interpolation algorithm satisfies some immediate properties, which are identified by \cite{BarraultNguyenMadayPatera2004} and \cite{MadayNguyenPateraPau2009} and summarised in the sequel:
\begin{description}
\item[Exact interpolation at magic points] For all functions $u\in \mathcal{U}$, the interpolation is exact at the magic points, in the sense that for every $m=1,\dots,M$
	\begin{equation}
	\label{eq:exact_interpolation}
		I_{m}(u)(z^\ast_j)=u(z^\ast_j)\qquad \text{for all } j\le m.
	\end{equation}
This property holds by construction of $q_m$. 
Note that $q_m(z^\ast_j)=0$ for $j<m$.
\item[Magic points as maxima] The basis function $q_m$ is maximal at the magic point $z^\ast_m$ i.e. 
	\begin{equation}
	\label{eq:maximum_point}
		q_m(z^\ast_m)=1=\sup_{z\in\Omega}|q_m(z)|.
	\end{equation}
\item[The matrix $B^M$ is invertible] By construction we get that the quadratic matrix $B^M\in\IR^{M\times M}$, introduced in \eqref{def-theta} as 
	\begin{equation*}
		B^M_{jm} = q_m(z^\ast_j)
	\end{equation*}
	is a lower triangular matrix with unity diagonal for all $j,m=1,\dots,M$. Its inverse thus exists.
\item[Coefficients of $I_m$ equal to those of $I_{m+1}$] The coefficients $\alpha^m_{j}=\alpha^m_j(u)$ of the interpolation  $I_m(u)=\sum_{j=1}^m \alpha^m_{j}q_j$ of $u$ do not depend on $m$, i.e. for all $i<m$ and $j\le i$ it holds that
	\begin{equation}
	\label{eq:coefficient_independent}
		\alpha^m_{j}=\alpha^i_{j}.
	\end{equation}
This can be seen from the triangular structure of the defining linear system for $\alpha^m=(\alpha^m_{j})_{j=1,\dots,m}$,
	\begin{equation}
		B^m\alpha^m=b^m\label{eq:solution_coefficients}
	\end{equation} 
with $b^m_j=u(z^\ast_j)$. By this representation we also get the linearity of $I_m$, for all $u,v\in \OU$,
	\begin{equation}
		I_m(u+v)=I_m(u)+I_m(v)\label{eq:linearity}.
	\end{equation}

\item[Idempotence] Let $1\le m \le M$. Since $I_m(v)=v$ for all $v\in\textrm{span}\{q_1,\dots,q_m\}$ we have for all $u\in\mathcal{U}$,
	\begin{equation}
	\label{eq:interpolation_idempotency}
		I_m(I_{m-1}(u))=I_{m-1}(u).
	\end{equation}
\end{description}

\subsection{Proof of Proposition \ref{prop-conv}}\label{sec-magicproof}

We give a detailed version of the proof of Theorem 2.4 in \cite{MadayNguyenPateraPau2009} with some minor deviations.
\begin{proof}
Recall that $I_{m-1}(u_m)=\sum_{j=1}^{m-1}\alpha^{m-1}_{j}q_j$, where $\alpha^{m-1}_{j}=\alpha^{m-1}_{j}(u_m)$. In order to get an upper bound for the absolute values of the coefficients $\alpha^{m-1}_{j}$, $j=1,\dots,m-1$, we use the triangular structure of the linear system\tild\eqref{eq:solution_coefficients} to obtain
\begin{align*}
\alpha^{m-1}_{j}=q_m(z^\ast_j)-\sum_{i=1}^{j-1}\alpha^{m-1}_{i}q_i(z^\ast_j).
\end{align*}
We then get $|\alpha^{m-1}_{1}|\le 1$ and for $j=1,\dots,m-1$ 
we deduce
	\begin{equation}
	\label{eq:coefficient_bound}
		|\alpha^{m-1}_{j}|=\left|u_m(z^\ast_j)-\sum_{i=1}^{j-1}\alpha^{m-1}_{i}q_i(z^\ast_j)\right|\le 1+\sum_{i=1}^{j-1}2^{i-1}=2^{j-1}.
	\end{equation}	
Next, we define the residuals
	\begin{equation}
		r_m(z)=u_m(z)-I_{m-1}(u_m)(z)=u_m(z)-\sum_{j=1}^{m-1}\alpha^{m-1}_{j}q_j(z)\label{eq:residual}
	\end{equation}
for all $z\in\Omega$. Our assumption on the Kolmogorov $n$-width guarantees the existence of constants $c>0$ and $\alpha>\log(4)$ such that for every $n\in\nn$,
	\begin{equation*}
		d_n(\mathcal{U},\mathcal{X})=\inf_{\mathcal{U}_n\in\OEE(\OX,n)}\sup_{g\in\mathcal{U}}\inf_{f\in\mathcal{U}_n}\|g-f\| \le c \ee{-\alpha n}
,
	\end{equation*}
where $\OEE(\OX,n)$ is the set of all $n$ dimensional subspaces of $\OX=L^\infty(\Omega, \cc)$.
Thus, for $M\in\nn$ and every $c_1>c$ there exists a linear subspace $\mathcal{U}_{M-1}\subset\mathcal{X}$ such that for all $q_j$, $j<m$, there exists $v_j\in \OU_{M-1}$ such that
\begin{equation}\label{eq-c_1}
\|q_j-v_j\|_{\infty}\le c_1\ee{-\alpha (M-1)}=c_2\ee{-\alpha M}
\end{equation}
 with $c_2 = c_1\ee{\alpha}$. Moreover, there exists $v_m\in \OU_{M-1}$ with $\|u_m-v_m\|_{\infty}\le c_2\ee{-\alpha M}$.
Setting $w_m:=v_m - \sum_{j=1}^{m-1}\alpha^{m-1}_{j}v_j$ and using the upper bounds on the absolute values of the coefficients $\alpha^{m-1}_{j}$ from inequality \eqref{eq:coefficient_bound} we get
\begin{align*}
\|r_m-w_m\|_{\infty}&\le c_2\ee{-\alpha M}\left(1+\sum_{j=1}^{m-1}|\alpha^{m-1}_{j}|\right)\le c_2\ee{-\alpha M}\left(1+\sum_{j=1}^{m-1}2^{j-1}\right)\\
&=c_2\ee{-\alpha M}2^{m-1}.
\end{align*}
By construction, dim$(\OU_{M-1})=M-1$, and thus we can find $\beta_1,\dots,\beta_M$ such that
\[
\sum_{m=1}^M\beta_m w_m = 0,
\]
where $|\beta_m|\le 1$ for all $m=1,\dots,M$ and $\beta_o=1$ for some $1\le o\le M$. This allows us to conclude that
\[
\left\|\sum_{m=1}^{M}\beta_m r_m\right\|_{\infty}=\left\|\sum_{m=1}^{M}\beta_m(r_m-w_m)\right\|_{\infty}\le c_2\ee{-\alpha M}M2^{M-1}.
\]
From equation \eqref{eq:exact_interpolation} we know that interpolation at the magic points is exact and hence $r_m(z^\ast_j)=u_m(z^\ast_j)-I_{m-1}(u_m)(z^\ast_j)=0$ for $j<m$ and thus
\[
|\beta_1||r_1(z^\ast_1)|=\left|\sum_{m=1}^{M}\beta_m r_m(z^\ast_1)\right|\le c_2\ee{-\alpha M}M2^{M-1}.
\]
Iteratively, we find
\begin{align*}
|\beta_m||r_m(z^\ast_m)|&= \left|\beta_mr_m(z^\ast_m)+\sum_{j=1}^{m-1}\beta_j r_j(z^\ast_m)-\sum_{j=1}^{m-1}\beta_j r_j(z^\ast_m)\right|\\
&\le\left|\sum_{j=1}^{M}\beta_jr_j(z^\ast_m)\right|+\left|\sum_{j=1}^{m-1}\beta_jr_j(z^\ast_m)\right|\\
&\le c_2\ee{-\alpha M}M2^{M-1}\left(1+\sum_{j=1}^{m-1}2^{j-1}\right)=2^{m-1}c_2\ee{-\alpha M}M2^{M-1}.
\end{align*}
Using that $\beta_o=1$ in the previous inequality, we immediately get
\begin{equation}
|r_o(z^\ast_o)|\le 2^{o-1}c_2\ee{-\alpha M}M2^{M-1}\label{eq:bound_o}.
\end{equation}
In the sequel we derive a bound for $\|r_M\|_\infty$ in the case $M>o$. For all $u\in\mathcal{U}$ we conclude that
\begin{align*}
\|u-I_m(u)\|_{\infty}&\le \|u-I_{m-1}(u)\|_{\infty}+\|I_m(u-I_{m-1}(u))\|_{\infty}\\
&=\|u-I_{m-1}(u)\|_{\infty}+\|I_m(u)-I_{m-1}(u)\|_{\infty},
\end{align*}
where we used identities \eqref{eq:linearity} and  \eqref{eq:interpolation_idempotency}.

Equation \eqref{eq:coefficient_independent} shows that $\alpha^m_{j}=\alpha^m_j(u)$ is independent of $m$ and thus $I_m(u)-I_{m-1}(u)=\alpha^m_{m}q_m$. By equation \eqref{eq:maximum_point} we know that $q_m$ is maximal at $z^\ast_m$ and together with equation \eqref{eq:exact_interpolation} we thus get
\begin{align*}
\|I_m(u)-I_{m-1}(u)\|_{\infty}&=\left|u(z^\ast_m)-I_{m-1}(u)(z^\ast_m)\right|\\
&\le\sup_{z\in\Omega}\left|u(z)-I_{m-1}(u)(z)\right|=\|u-I_{m-1}(u)\|_{\infty}.
\end{align*}
The last two results iteratively yield for $j\le m$,
\[
\|u-I_m(u)\|_{\infty}\le2^j\|u-I_{m-j}(u)\|_{\infty}.
\]
Finally, with inequality \eqref{eq:bound_o} we conclude
\[
\|u-I_M(u)\|_{\infty}\le 2^{M-o}\|u-I_{o}(u)\|_{\infty}\le 2^{M-o}\|r_o\|_{\infty}\le c_22^{2M-2}M\ee{-\alpha M}
\]
and this proves the claim.
\end{proof}

\section{Proof of Theorem \ref{theo:MFTExpConvEU}}\label{sec-proofExpo}

The following proof is taken from \cite{GassGlau2015}.

\begin{proof}
In principle, we exploit the analyticity property of $H_1$ from condition\tild\ref{cond:MFTExpConvEU} to estimate the Kolmogorov $n$-width of the set $\OU$. This can conveniently be achieved by inserting an example of an interpolation method that is equipped with exact error bounds. We choose Chebyshev polynomial interpolation for this task. For polynomials of degree $N\in\nn$, the Chebyshev nodes are given by ${z}_k = \cos\left(\pi\frac{2k+1}{2N+2}\right)$ for $k=0,\ldots,N$ and the basis functions are defined as
\begin{align}\label{eq-Tjp}
  T_j(z) := \cos\big(j \arccos(z)\big)\qquad\text{for $z\in [-1,1]$ and $0\le j\le N$.}
   \end{align}
For fixed $p\in \OP$, the Chebyshev interpolation of $H_1(p,\cdot)$ with Chebyshev polynomials of degree $N$ is of the form
 \begin{align}\label{eq-ChebInter1dim}
I^\text{Cheby}_N(H_1(p,\cdot))(z)&:= \sum_{j=0}^N c_j T_j(z)
 \end{align}
 with coefficients 
 $c_j:=\frac{2^{\1_{j>0}}}{N+1}\sum_{k=0}^N H_1(p,z_k)\cos\big(j\pi\frac{2k +1}{2N+2}\big)$, $j=0,\ldots,N$. From Theorem\tild 8.2 in \cite{Trefethen2013} we obtain the explicit error bound 
\begin{equation}\label{Cheby-error}
\sup_{p\in\OP}\big\| H_1(p,\cdot) - I^\text{Cheby}_N\big(H_1(p,\cdot)\big) \big\|_{\infty} \le C_1(H_1) \varrho^{-N} 
\end{equation} 
with constant $C_1(H_1):=\frac{4}{\varrho-1}\max_{(p,z)\in \OP\times B(\Omega,\varrho)} \big|H_1(p,z)\big|$.

The Chebyshev interpolation of the family of functions $H_1(p,\cdot)$, $p\in\OP$ induces an approximation of the family of functions $h(p,\cdot)$, $p\in\OP$, along with an $N$-dimensional function space $\mathcal{U}_N$, simply by setting
\begin{equation}
I^\text{Kolm}_N\big(h(p,\cdot) \big) (z) := I^\text{Cheby}_N\big(H_1(p,\cdot)\big)(z) H_2(z)
\end{equation} 
 for all $z\in\Omega$ and $p\in\OP$. The approximation $I^\text{Kolm}_N$ inherits the error bound
 \begin{equation}\label{IK-error}
\sup_{(p,z)\in\OP\times\Omega}\big| h(p,z) - I^\text{Kolm}_N\big(h(p,\cdot)\big)(z) \big| \le C_2 \varrho^{-N} 
\end{equation} 
with constant $C_2:=C_1(H_1)\max_{z\in\Omega}|H_2(z)|$  from \eqref{Cheby-error}. From \eqref{IK-error}, we obtain an upper bound for the Kolmogorov $n$-width so that we can apply the general convergence result from 
Theorem 2.4 in \cite{MadayNguyenPateraPau2009}. Consulting their proof, respectively inserting \eqref{IK-error} in inequality\tild\eqref{eq-c_1} in  Appendix \ref{sec-magicproof},
we realize that
\begin{align*}
\sup_{p\in\OP}\big\|h(p,\cdot)-I_M(h)(p,\cdot)\big\|_{\infty} &\le CM(\varrho/4)^{-M}
\end{align*} 
with $C=\frac{C_1\varrho}{4}$. The estimation of the error of the Magic Point Integration now follows by integrating with respect to $z$.
\end{proof}

\section{Truncation Error in Fourier Pricing}
\label{sec:MFTTruncationError}

We introduce the following condition
that is satisfied for a large class of models and payout profiles:

\textit{For every $N>0$, there exist constants $\alpha,C_1,C_2,m>0$ such that uniformly for every $(K,T,q)\in\OP=\OK\times\OT\times\OQ$,
\begin{align}\label{Para}
\Re\big(\log(\varphi_{T,q}(\xi+i\eta))\big) &\le - C_1|\xi|^\alpha \,&\text{for all $|\xi|>N$},\tag{G{\aa}rd}\\
\Big|\widehat{f_K}(\xi + i\eta) \Big|&\le C_2|\xi|^m \quad &\text{for all $|\xi|>N$}.\tag{Poly}
\end{align}
}
Virtually for every payoff profile $f_K$ the generalized Fourier transform \text{$\widehat{f_K}(\cdot \!+\! i\eta)$} exists for some $\eta\in\rrd$ and decays polynomially, uniformly in a reasonably large set of parameters $K$.
Condition (G{\aa}rd) already appears in another context, where it implies that the related bilinear form satisfies a so-called G{\aa}rding condition with respect to fractional Sobolev spaces of order $\alpha$. This helps to classify the solution spaces of related weak solutions to associated Kolmogorov equations. For a proof of this implication as well as for numerous examples of classes of (time-inhomogeneous) L\'evy processes satisfying the condition we refer to \cite{Glau2015a} for the case $\eta=0$ and to \cite{Glau2015b} for $\eta\neq0$. 
The following proposition is immediate.
\begin{lemma}[Truncation error]
Assume for every $(K,T,q)\in \OP$, \emph{(G{\aa}rd)} and \emph{(Poly)} and let $\Omega\subset \rr_+\times\rrd + i\eta$ and denote $|\Omega|$ the diameter of the largest ball centered in the origin that is contained in\tild $\Omega$. For every $\beta<\alpha$ there exists a constant $c>0$ such that uniformly for every $(K,T,q)\in\OP$,
\begin{align*}
\Bigg|\int_{\rr_+\times\rrd + i\eta } \!\! \Re\Big(\widehat{f_K}(-z) \varphi_{T,q}(z) \Big)\dd z - \int_{\Omega}\! \Re\Big( \widehat{f_K}(-z) \varphi_{T,q}(z) \Big) \dd z  \Bigg| \le c \ee{-\beta|\Omega|}.
\end{align*}
\end{lemma}

\bibliographystyle{elsarticle-harv}
  \bibliography{LiteraturFourierRB}

\end{document}